\documentclass[fleqn,usenatbib]{mnras}

\usepackage{newtxtext,newtxmath}

\usepackage[T1]{fontenc}

\DeclareRobustCommand{\VAN}[3]{#2}
\let\VANthebibliography\thebibliography
\def\thebibliography{\DeclareRobustCommand{\VAN}[3]{##3}\VANthebibliography}

\usepackage{graphicx}	
\usepackage{amsmath}	
\usepackage{amssymb}	
\usepackage{color}
\usepackage{soul}
\usepackage{siunitx}
\usepackage{threeparttable}
\usepackage{subfigure} 
\makeatletter
\renewcommand{\@thesubfigure}{\hskip\subfiglabelskip}
\makeatother

\title[SN 2018hfm: A Low-Energy Type II Supernova]{SN 2018hfm : A Low-Energy Type II Supernova with Prominent Signatures of Circumstellar Interaction and Dust Formation}

\author[X. Zhang et al.]{
Xinghan Zhang$^{1}$\thanks{E-mail: xh-zhang17@mails.tsinghua.edu.cn},
Xiaofeng Wang$^{1,2}$\thanks{E-mail: wang\_xf@mail.tsinghua.edu.cn},
Hanna Sai$^{1}$,
Maria Niculescu-Duvaz$^{3}$,
Alexei V. Filippenko$^{4,5}$,
\newauthor
WeiKang Zheng$^{4}$,
T. G. Brink$^{4}$,
Han Lin$^{1}$,
Jicheng Zhang$^{1}$,
Yongzhi Cai$^{1}$,
Jun Mo$^{1}$,
Jujia Zhang$^{6,7,8}$,
\newauthor
E. Baron$^{9}$,
J. M. DerKacy$^{9}$,
F. Huang$^{10}$,
T.-M. Zhang$^{11,12}$
\\
$^{1}$Physics Department and Tsinghua Centre for Astrophysics (THCA), Tsinghua University, Beijing 100084, China\\
$^{2}$Bejing Planetarium, Beijing Academy of Science and Technology, Beijing 100044, China\\
$^{3}$Department of Physics and Astronomy, University College London, Gower St., London WC1E 6BT, UK\\
$^{4}$Department of Astronomy, University of California, Berkeley, CA 94720-3411, USA\\
$^{5}$Miller Institute for Basic Research in Science, Univerisity of California, Berkeley, CA 94720, USA\\
$^{6}$Yunnan Observatories, Chinese Academy of Sciences, Kunming 650011, China \\
$^{7}$Key Laboratory for the Structure and Evolution of Celestial Objects, Chinese Academy of Sciences, Kunming 650216, China\\
$^{8}$Center for Astronomical Mega-Science, Chinese Academy of Sciences, 20A Datun Road, Chaoyang District, Beijing 100012, China\\
$^{9}$Homer L. Dodge Department of Physics and Astronomy, University of Oklahoma, Norman, OK 73019, USA\\
$^{10}$Department of Astronomy, Shanghai Jiao Tong University, Shanghai, 200240, China\\
$^{11}$Key Laboratory of Optical Astronomy, National Astronomical Observatories, Chinese Academy of Sciences, Beijing 10101, China\\
$^{12}$School of Astronomy and Space Science University of Chinese Academy of Sciences, Beijing 101408, China
}

\date{Accepted 2021-10-13. Received YYY; in original form ZZZ}

\pubyear{2020}

\begin{document}
\label{firstpage}
\pagerange{\pageref{firstpage}--\pageref{lastpage}}
\maketitle

\begin{abstract}
We present multiband optical photometric and spectroscopic observations of an unusual Type II supernova, SN 2018hfm, which exploded in the nearby ($d \approx 34.67$\,Mpc) dwarf galaxy PGC 1297331 with a very low star-formation rate (0.0270\,M$_{\odot}$\,yr$^{-1}$) and a subsolar metallicity environment ($\sim 0.5$\,Z$_{\odot}$). The $V$-band light curve of SN 2018hfm reaches a peak with value of $-18.69\pm0.64$\,mag, followed by a fast decline ($4.42\pm0.13$\,mag\,(100\,d)$^{-1}$). After about 50 days, it is found to experience a large flux drop ($\sim 3.0$\,mag in $V$), and then enters into an unusually faint tail, which indicates a relatively small amount of $^{56}$Ni synthesised during the explosion. From the bolometric light curve, SN 2018hfm is estimated to have low ejecta mass ($\sim1.3$\,M$_{\odot}$) and low 
explosion energy ($\sim 10^{50}$\,erg) compared with typical SNe~II. The photospheric spectra of SN 2018hfm are similar to those of other SNe~II, with P~Cygni profiles of the Balmer series and metal lines, while at late 
phases the spectra are characterised by box-like profiles of H$\alpha$ emission, suggesting significant interaction between the SN ejecta and circumstellar matter. These box-like emission features are found to show increasing asymmetry with time, with the red-side component becoming gradually weaker, indicating that dust is continuously formed in the ejecta. Based on the dust-estimation tool \textsc{damocles}, we find that the dust increases from $\sim 10^{-6}$\,M$_{\odot}$ to $10^{-4}$--$10^{-3}$\,M$_{\odot}$ between +66.7\,d and +389.4\,d after explosion.
\end{abstract}




\section{Introduction}
Type II supernovae (SNe~II) are explosions resulting from core collapse in massive stars ($\ge 8$\,M$_{\odot}$) with manifest hydrogen features in 
their optical spectra. They can be further divided into the following subclasses: SNe~IIP, characterised by a long plateau ($\sim 100$\,d) in the light curve (LC) followed by a rapid drop; SNe~IIL, whose LC declines linearly (in magnitudes) after the peak; SNe~IIn, where ``n'' denotes relatively narrow emission lines formed from the interaction between SN ejecta and circumstellar matter (CSM); and SNe~IIb, those showing similar spectra to SNe~IIP/L near maximum brightness but resembling SNe~Ib in the following weeks \citep{Filippenko1997,Gal-Yam2017}. Recent large-sample studies favour that SNe~IIP and IIL actually constitute a continuous distribution rather than a bimodal one, with more-luminous SNe tending to decline faster after the peak \citep{A14,V16}. Additionally, \citet{Valenti2015} proposed that all supernovae that were previously classified as SNe IIL actually have a sudden flux drop before their light curves enter into the tail phase as long as they have been observed for enough long time. In the continuous distribution hypothesis, light curves of SNe II all have four-stage evolution: a rising phase, a plateau-like phase\footnote{``Plateau'' is a terminology from SNe~IIP, referring to the level part of their light curves. As we accept the continuous distribution hypothesis, we generalize the term to refer to the phase between the peak of light curve and the sudden flux drop, despite of any decline trend at this phase. In the following text, the terms, ``plateau-like phase'', ``plateau phase'' or ``plateau'', all refer to this generalized meaning.}, a sudden flux drop\footnote{The ``sudden flux drop'' is also called as ``transition phase''.} and a tail phase. Note that SNe~IIn may not represent an intrinsically distinct SN type but an external phenomenon produced by circumstellar interaction (CSI; \citealt{Smith2017_handbookSN,Schlegel1990}), for which the true SN component is concealed by photons from the interaction.\\

In addition to SNe~IIn, signatures of CSI have been observed in many SNe IIP, IIL or IIb, such as SN 2007od \citep{Andrews2010,Inserra2011}, SN 2016gfy \citep{Singh2019}, SN 2004et \citep{Kotak2009}, SN 2017eaw \citep{Rui2019, Weil2020}, and SN 1993J \citep{Patat1995, Matheson2000a, Matheson2000b}, usually manifested as broad, boxy emission lines of hydrogen, sometimes accompanied by forbidden lines of oxygen and calcium. When the SN ejecta collide with the CSM, forward and reverse shocks are created, between which a cold dense shell exists.  The energy, mainly from the reverse shock, heats and ionises the surrounding material to form shell-like emission regions \citep{CF94}. Emission from an optically thin and homologously expanding shell forms a box-like profile in the spectrum, with the velocity at zero intensity corresponding to the outer boundary of the shell, and the maximum velocity at the flat top linked to the inner boundary of the shell \citep{Patat1995, Jerkstrand2017_handbookSN, BevanPhDT}. Since the CSI lies at the very outer layers of the SN ejecta, the box-like profiles are usually very broad.\\

Some high-redshift galaxies have been found to contain large amounts of dust \citep{Bertoldi2003}, and core-collapse SNe (CCSNe) are thought to be 
potential dust factories \citep{Kozasa1991}. This indicates that each CCSN should roughly contribute 0.1--1.0\,M$_{\odot}$ of dust if they are responsible for all of the dust observed in a galaxy. However, this is not seen in observations, which show that only $10^{-5}$--$10^{-4}$\,M$_{\odot}$ of dust can be produced by a typical SN~II within 1000\,d after explosion \citep{Kotak2009,Meikle2007,Andrews2010}.\\

Normally, dust formation leaves three signatures --- a drop in optical brightness due to extinction by dust, an infrared excess coming from radiation reemitted by dust, and red-blue asymmetry of emission-line profiles caused by attenuation from dust \citep{Andrews2010}. Based on modelling the red-blue asymmetric profiles observed in some emission lines, \citet{Bevan2016} and \citet{Bevan2019,Bevan2020} found that dust is continuously formed in the SN ejecta, which could alleviate the tension between theoretical expectations and observations. Nevertheless, such an analysis is only limited to a few cases (such as SN 1987A, SN 2005ip, and SN 2010jl), so more samples are needed to further verify this possibility. SN 2018hfm is an example showing box-like emission profiles with developing red-blue 
asymmetry, providing us a rare opportunity to study the CSM environment and dust formation in SN ejecta.\\

Our observations and data reduction are described in Section~\ref{sec:obs}. We estimate the extinction and distance to SN 2018hfm in Section~\ref{sec:extinction}. Section~\ref{sec:host} investigates properties of the host galaxy. The SN's photometric and spectroscopic evolution are presented 
in Section~\ref{sec:phot} and \ref{sec:spec}, respectively. We discuss the CSI, progenitor properties, and dust formation of SN 2018hfm in Section~\ref{sec:discussion}. Section~\ref{sec:conclusion} provides a summary of 
our work.

\section{Observations and data reduction}
\label{sec:obs}
\begin{figure*}
   \includegraphics[width=2\columnwidth]{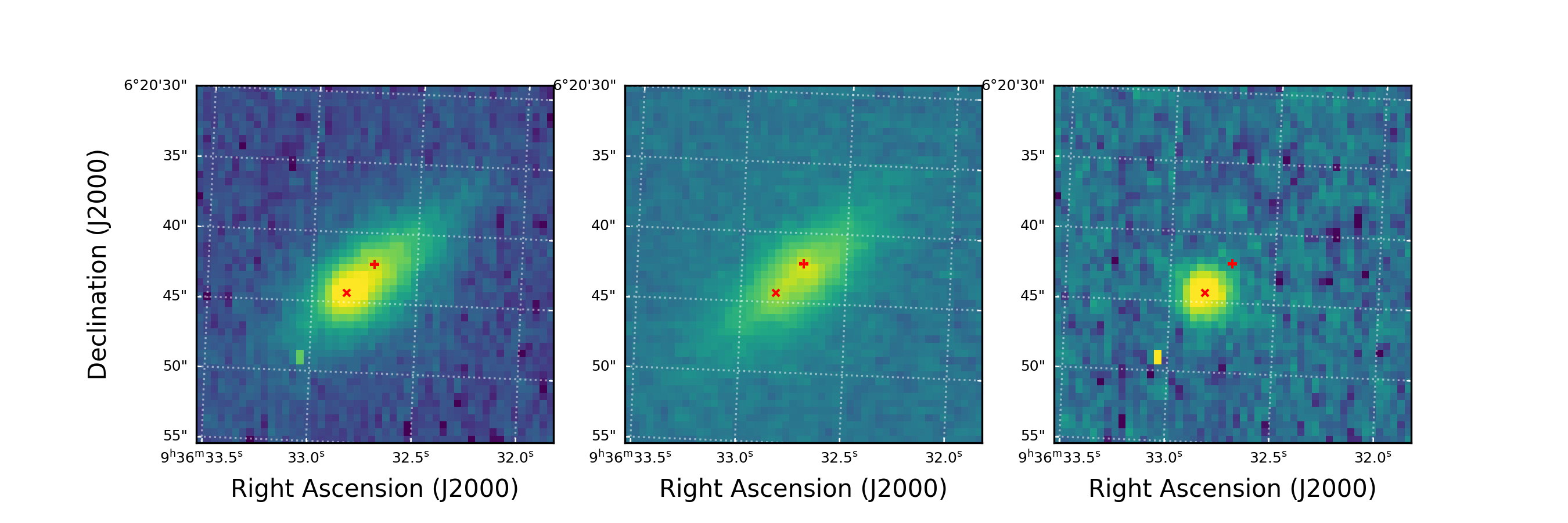}
   \caption{\textit{Left}: An $r$-band image of SN 2018hfm and its host galaxy taken on 14 Dec. 2018. \textit{Middle}: A very late-time image taken when SN 2018hfm had faded away. \textit{Right}: The difference image obtained by subtracting the middle image from the left one. The red cross in the image denotes the location of SN 2018hfm, while the red plus sign marks the centre of the host galaxy.}
   \label{fig:image}
\end{figure*}

SN 2018hfm was discovered and first reported on 9 Oct. 2018 at 15:07:12.00 (UT dates are used throughout this paper; MJD = 58400.630) by the ASASSN (All-Sky Automated Survey for Supernovae) team \citep{2018TNSTR1539....1S}, with J2000 coordinates of $\alpha = 09^{\rm h}36^{\rm m}08.736^{\rm s}$ and $\delta = +06^\circ15'27.08''$. As shown in Figure \ref{fig:image}, this source is located $\sim2.8\arcsec$ northeast from the centre of the host galaxy, PGC 1297331. Two days after the discovery, a blue featureless spectrum was obtained by \citet{2018TNSCR1554....1Z}, and another spectrum taken one day later shows a shallow absorption due to \ion{He}{i} $\lambda$5876, with an expansion velocity of $\sim 7500$\,km\,s$^{-1}$ \citep{2018TNSCR2156....1N}. However, neither of these two spectra provides a conclusive classification of SN 2018hfm. Later follow-up spectra identify this transient as a Type II SN with some peculiar characteristics, such as high-velocity features and a box-like profile of H$\alpha$ emission. The host galaxy of SN 2018hfm has redshift $z = 0.008$ according to NED \citep{2018TNSCR2156....1N}.

\subsection{Photometry}
\subsubsection{ASASSN and Nickel data}

Our multiband photometric observations (see Sec.\ref{sec:TNTdata}) began on 24 Nov. 2018 (MJD = 58446), $\sim 45$\,d after discovery. We supplement these data with earlier observations from ASASSN obtained via the \texttt{Sky Patrol}\footnote{\url{https://asas-sn.osu.edu}} website \citep{Shappee2014,Kochanek2017}. After providing coordinates and dates to the website, real-time aperture photometry with zeropoints calibrated using the APASS catalog was automatically applied to the unpublished $V$/$g$-band 
images, and we obtained the measured fluxes.\\

Note that the standard method of measuring the SN flux is to perform photometry on the residual image after a reference image is subtracted. However, since raw images were unavailable to us, we adopted an alternative way to estimate the flux of SN 2018hfm using the ASASSN data. We first estimated contamination from the host galaxy by performing real-time aperture 
photometry on images taken within 400\,d before explosion and then calculated the median ($f_{\rm bkg}$) and standard deviation ($\sigma_{\rm bkg}$) of the obtained flux. Then we focused on the flux observed around the discovery time of SN 2018hfm, selecting points higher than $f_{\rm bkg} + 
3\sigma_{\rm bkd}$ as being definitely detected signals. After subtracting $f_{\rm bkg}$, we obtained the ``pure'' SN flux; the results are listed 
in Table~\ref{tab:ASNphoto}.\\

\begin{table*}
	\centering
	\caption{SN 2018hfm $V/g$-band ASASSN photometry}
	\label{tab:ASNphoto}
	\begin{threeparttable}
	\begin{tabular}{cccc} 
		\hline
		MJD     & $V$        & MJD     & $g$      \\
		\hline
		58392.6 & >17.97      & 58401.5 & 15.04(03)  \\
		58400.6 &  14.97(04)  & 58404.5 & 15.10(03)  \\
		58414.6 &  15.70(08)  & 58426.5 & 16.44(08)  \\
		58417.6 &  15.87(09)  & 58437.5 & 17.26(12) \\
		58430.6 &  16.42(08)  & 58439.3 & 17.03(10) \\
		58432.6 &  16.59(09)  & 58443.5 & 17.43(17) \\
		58436.5 &  16.42(08)  &         &           \\
		58442.5 &  16.70(15)  &         &           \\
		\hline
	\end{tabular}
	      \begin{tablenotes}
           \footnotesize
           \item Note: numbers in parentheses are uncertainties in units of 0.01\,mag.
        \end{tablenotes}
	\end{threeparttable}
\end{table*}


Images of SN 2018hfm were also obtained by the 1\,m Nickel telescope (Nickel, hereafter) at Lick Observatory in the $BVRI$ bands on two nights (MJD = 58419.5 and 58449.5). The data were reduced using an image-reduction pipeline \citep{Ganeshalingam2010, Stahl2019}. Point-spread-function (PSF) photometry using DAOPHOT \citep{Stetson1987} was performed after a template image was subtracted. Several nearby stars were chosen from the Pan-STARRS1\footnote{\url{http://archive.stsci.edu/panstarrs/search.php}} catalog to help calibrate the flux of SN 2018hfm. The results are listed in Table~\ref{tab:NickelTphoto}. We find that the Nickel data are quite consistent with the ASASSN data, thus enhancing the reliability of the method we used in estimating the flux of SN 2018hfm from the ASASSN 
\texttt{Sky Patrol} site.

\begin{table*}
	\centering
	\caption{SN 2018hfm Photometry from Lick/Nickel}
	\label{tab:NickelTphoto}
	\begin{threeparttable}
	\begin{tabular}{ccccc} 
		\hline
		MJD     & $B$        & $V$     & $R$     & $I$ \\
		\hline
		58419.488 & 16.315(025) & 15.856(016) & 15.486(015) & 15.213(018)\\
		58449.486 & 18.507(145)	& 17.920(109) & 16.753(070)	& 16.889(115)\\
		\hline
	\end{tabular}
	    \begin{tablenotes}
	       \item Note: numbers in parentheses are uncertainties in units of 0.001\,mag.
	    \end{tablenotes}
\end{threeparttable}
\end{table*}

\subsubsection{TNT data}
\label{sec:TNTdata}

Multiband photometric data were also collected with the Tsinghua-NAOC 0.8\,m Telescope (TNT, hereafter; \citealt{Huang2012}) at Xinglong Observatory in the Johnson-Cousins $BV$ and SDSS (Sloan Digital Sky Survey) $gri$ bands. All CCD images were preprocessed using standard \textsc{IRAF}\footnote{\textsc{IRAF} is distributed by the National Optical Astronomy Observatories, which are operated by the Association of Universities for Research in Astronomy, Inc., under cooperative agreement with the U.S. National Science Foundation (NSF)} routines, including corrections for bias, flat field, and removal of cosmic rays. We performed astrometric calibration 
on all of our images using the \texttt{Astrometry.net} software \citep{2012ascl.soft08001L}. Then all of the images were reprojected into coordinates of the template using a Python Package \texttt{reproject}\footnote{\url{https://reproject.readthedocs.io/en/stable/}}. The template was taken on 6 May 2019 (MJD = 58609), 208\,d after the discovery, when the SN was dimmer than the TNT detection limit. We subtracted the template from all of the images using \textsc{hotpants} \citep{2015ascl.soft04004B} and performed aperture photometry on the residual images to obtain instrumental magnitudes through \texttt{AstroImageJ}\footnote{\url{https://www.astro.louisville.edu/software/astroimagej/}} \citep{2013ascl.soft09001C}. We 
extracted sources in the template image and performed photometry on them using a bundled program \texttt{image2xy} of the tool \texttt{Astrometry.net}. After cross-matching these sources to the APASS and SDSS catalogs, we obtained the flux-calibrated magnitudes of SN 2018hfm. \\

\begin{figure*}
	\includegraphics[width=2\columnwidth]{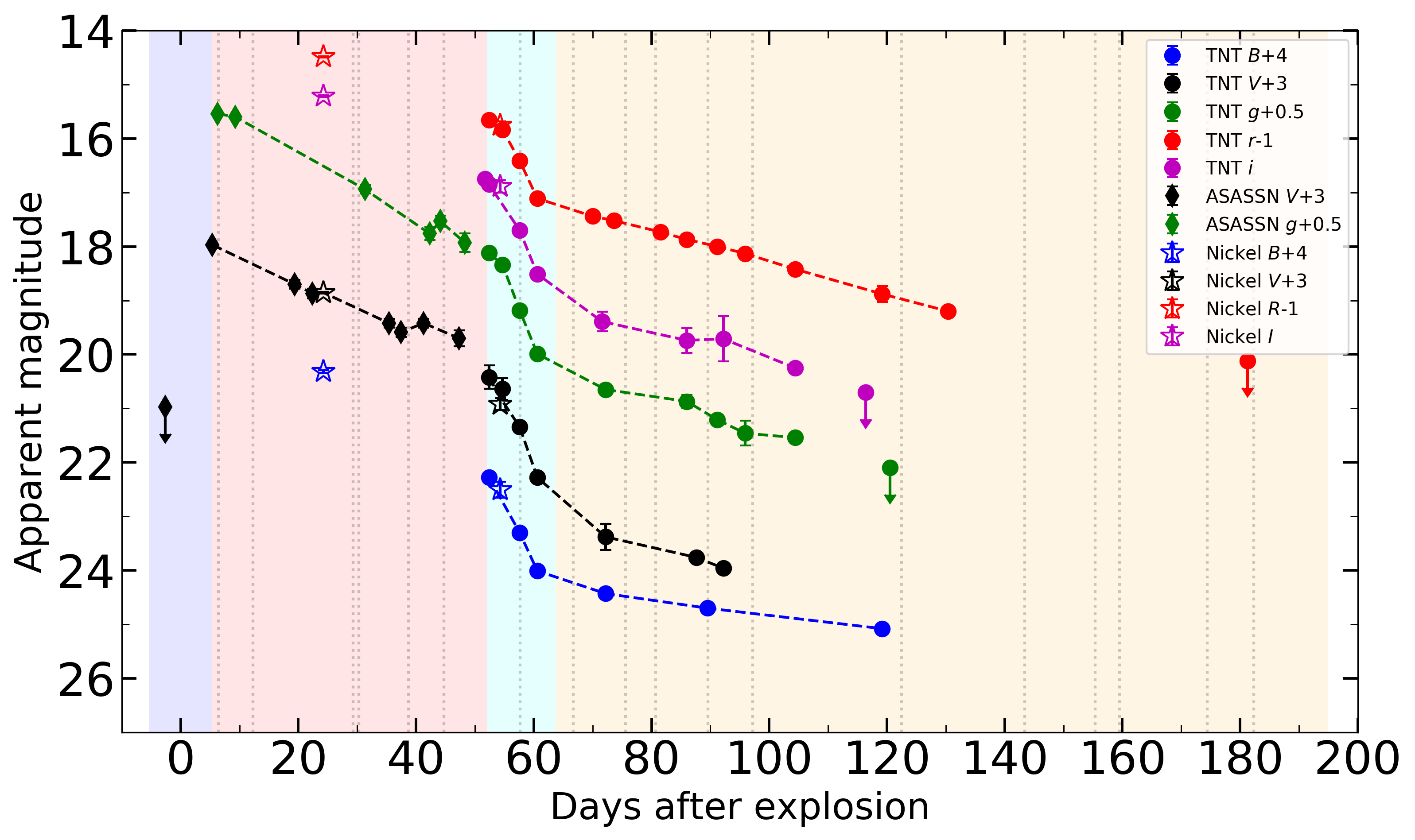}
    \caption{Multiband light curves of SN 2018hfm obtained with the TNT (filled circles), the Lick/Nickel (hollow stars), and from the ASASSN \texttt{Sky Patrol} (diamonds). Data points are shifted vertically, if necessary, for clarity. An upper limit is denoted by a downward arrow. Vertical 
dashed lines denote spectral epochs and shaded regions show four evolution phases: rising, plateau-like, transitional, and tail (from left to right).}
    \label{fig:LC}
\end{figure*}

All of the photometric data, spanning from 5.4\,d to 116.4\,d after the explosion, are displayed in Table~\ref{tab:TNTphoto} and Figure~\ref{fig:LC}.

\begin{table*}
	\centering
	\caption{SN 2018hfm photometry from the TNT}
	\label{tab:TNTphoto}
	\begin{threeparttable}
	\begin{tabular}{cccccc} 
		\hline
		MJD     & $B$      & $V$       & $g$       & $r$       & $i$     \\
		\hline
		58446.9 & --       & --        & --        & --        & 16.75(01) \\
		58447.6 & 18.28(02) & 17.42(22) & 17.62(01)  & 16.66(03)  & 16.85(01) \\
		58449.9 & --       & 17.64(20)  & 17.84(01)  & 16.84(01)  & --       \\
		58452.8 & 19.30(01) & 18.34(01)  & 18.68(01)  & 17.41(01)  & 17.70(01) \\
		58455.8 & 20.01(01) & 19.28(01)  & 19.49(01)  & 18.11(01)  & 18.51(01) \\
		58465.2 & --       & --        & --        & 18.44(01)  & --       \\
		58466.8 & --       & --        & --        & --        & 19.39(18)\\
		58467.4 & 20.43(08) & 20.38(24) & 20.15(10) & --        & --       \\
		58468.8 & --       & --        & --        & 18.52(02)  & --       \\
		58476.7 & --       & --        & --        & 18.73(01)  & --       \\
		58481.2 & --       & --        & 20.37(12) & 18.87(10) & 19.74(23) \\
		58482.8 & --       & 20.76(01)  & --        & --        & --        \\
		58484.7 & 20.70(07) & --        & --        & --        & --        \\
		58486.4 & --       & --        & 20.71(02)  & 19.00(06)  & --        \\
		58487.4 & --       & 20.96(01)  & --        & --        & 19.71(42) \\
		58491.1 & --       & --        & 20.96(23) & 19.13(02)  & --        \\
		58499.6 & --       & --        & 21.04(01)  & 19.42(09)  & 20.25(01)  \\
		58511.6 & --       & --        & --        & --        & >20.70(01) \\
		58514.4 & 21.08(01) & --        & --        & 19.88(15) & --        \\
		58515.7 & --       & --        & >21.60    & --        & --        \\
		58525.6 & --       & --        & --        & 20.20(01)  & --        \\
		58576.5 & --       & --        & --        & >21.12    & --        \\
		\hline
	\end{tabular}
      \begin{tablenotes}
         \footnotesize
         \item Note: numbers in parentheses are uncertainties in units of 
0.01\,mag.
       \end{tablenotes}
\end{threeparttable}
\end{table*}

\subsection{Spectroscopy}
\begin{figure}
	\includegraphics[width=\columnwidth]{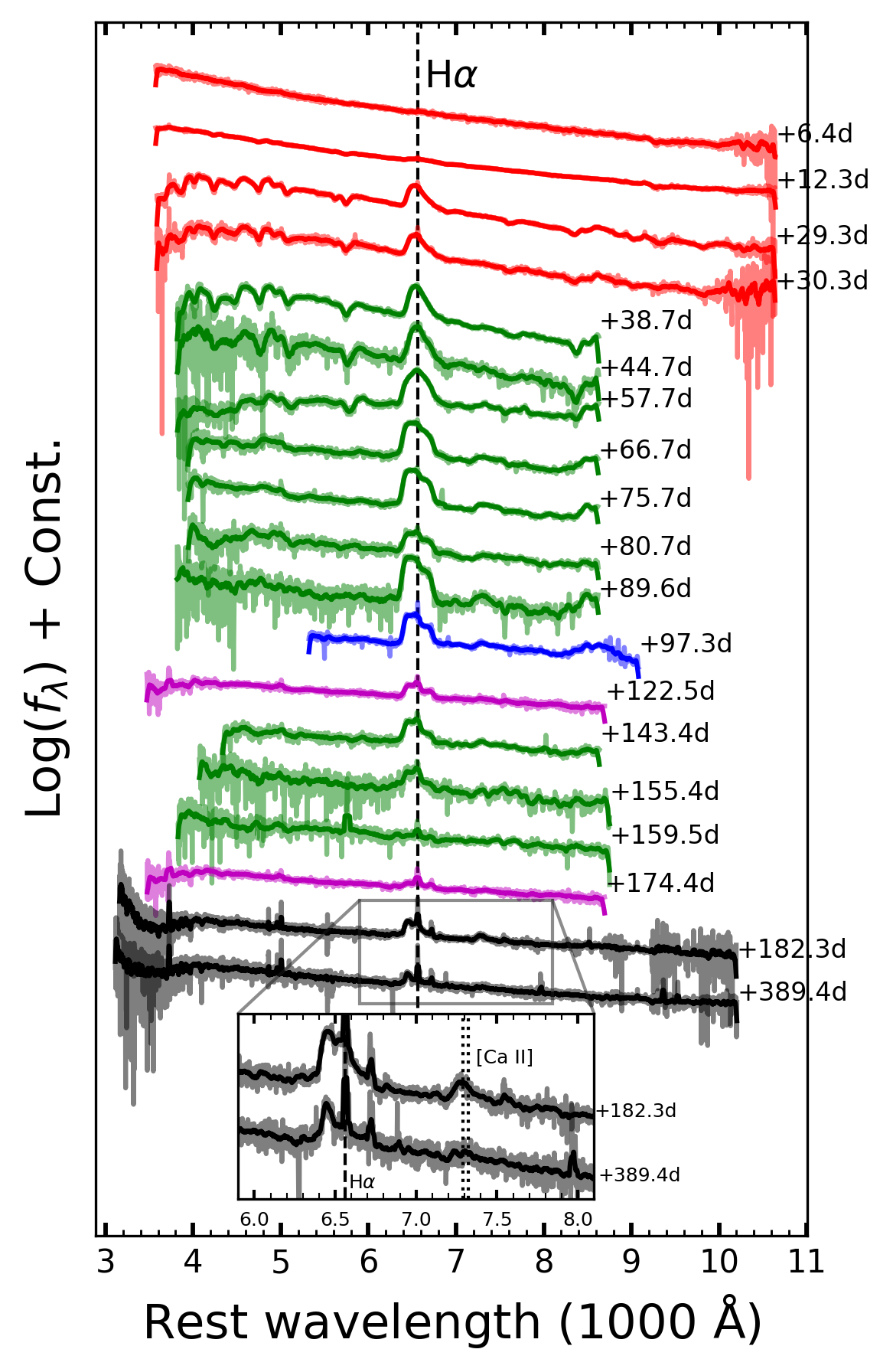}
    \caption{Optical spectroscopic evolution of SN 2018hfm. All of the spectra have been corrected for the redshift and extinction. They are smoothed (bold lines) and shifted vertically for better clarity. The phase after explosion (MJD = 58395.2) is marked on the right side of each spectrum. The last two spectra are partially zoomed-in for better display of the $\rm H\alpha$ and [\ion{Ca}{ii}] emission lines. Spectra from different 
instruments are shown in different colours, with red representing spectra 
from Lick, green from XLT, blue from APO, magenta from LJT, and black from Keck.}
    \label{fig:spectra}
\end{figure}

As shown in Figure~\ref{fig:spectra}, 19 spectra of SN 2018hfm were obtained: four with the Kast spectrograph \citep{Miller1993} on the 3\,m Shane telescope at Lick Observatory (Lick, hereafter), ten with the Beijing Faint Object Spectrograph and Camera (BFOSC) on the 2.16\,m telescope at Xinglong Observatory (XLT, hereafter), one with the Dual Imaging Spectrograph (DIS) on the 3.5\,m telescope at the Apache Point Observatory (APO, hereafter), two with the Yunnan Faint Object Spectrograph and Camera (YFOSC) on the Li-Jiang 2.4\,m telescope of Yunnan Astronomical Observatory (LJT, hereafter; \citealt{Fan2015}), and two (at late times) with the Low-Resolution Imaging Spectrometer (LRIS) on the Keck-I 10\,m telescope (Keck, hereafter; \citealt{Oke1995}). A Journal of spectroscopic observations is presented in Table~\ref{tab:spec}.\\

\begin{table*}
	\centering
	\caption{Journal of spectroscopic observations of SN 2018hfm}
	\label{tab:spec}
	\begin{threeparttable}
	\begin{tabular}{ccccccc} 
		\hline
		No.  & UT Date    & MJD   &Epoch (day)\tnote{*} & Exp. (s) & Telescope+Inst. & Range (\AA) \\
		\hline
		1    & 2018-10-10 & 58401.6 & 6.4   & $2 \times 300$  & Lick 3m+Kast & 3573 - 10642\\
		2    & 2018-10-16 & 58407.5 & 12.3  & 1500  & Lick 3m+Kast & 3573 - 10646\\
		3    & 2018-11-02 & 58424.5 & 29.3  & 1500  & Lick 3m+Kast & 3587 - 10640\\
		4    & 2018-11-03 & 58425.5 & 30.3  & 1800  & Lick 3m+Kast &  3585 - 10639\\
		5    & 2018-11-11 & 58433.9 & 38.7  & 3300  & XLT+BFOSC    &3817 - 8627\\
		6    & 2018-11-17 & 58439.9 & 44.7  & 3000  & XLT+BFOSC & 3820 - 8627\\
		7    & 2018-11-30 & 58452.9 & 57.7  & 3000  & XLT+BFOSC & 3821 - 8627 \\
		8    & 2018-12-09 & 58461.9 & 66.7  & 3300  &  XLT+BFOSC & 3941 - 8621 \\
		9    & 2018-12-18 & 58470.8 & 75.6  & 3300  & XLT+BFOSC  & 3941 - 8617 \\
		10   & 2018-12-23 & 58475.9 & 80.7  & 3300  & XLT+BFOSC  & 3945 - 8622 \\
		11   & 2019-01-01 & 58484.8 & 89.6  & 3600  & XLT+BFOSC  & 3818 - 8621 \\
		12   & 2019-01-09 & 58492.4 & 97.2  & 1800  & APO+DIS    & 5321 - 9081 \\
		13   & 2019-02-03 & 58517.7 & 122.5 & 2098  & LJT+YFOSC  & 3471 - 8696 \\
		14   & 2019-02-24 & 58538.6 & 143.4 & 3600  & XLT+BFOSC  & 4337 - 8634 \\
		15   & 2019-03-08 & 58550.6 & 155.4 & 3600  & XLT+BFOSC  & 4070 - 8752 \\
		16   & 2019-03-12 & 58554.7 & 159.5 & 3600  & XLT+BFOSC  & 3827 - 8751 \\
		17   & 2019-03-27 & 58569.6 & 174.4 & 2400  & LJT+YFOSC  & 3475 - 8694 \\
		18   & 2019-04-04 & 58577.5 & 182.3 & 607.5 & Keck I+LRIS & 3160 - 10196 \\
		19   & 2019-10-28 & 58784.6 & 389.4 & 900.0 & Keck I+LRIS & 3112 - 10204\\
		\hline
	\end{tabular}
	
       \begin{tablenotes}
           \footnotesize
             \item[*] The epoch is relative to the explosion date, MJD = 
58395.2.
       \end{tablenotes}
    \end{threeparttable}
\end{table*}

The spectra from XLT, LJT, and APO were reduced using standard \textsc{IRAF} routines including bias correction, flat fielding, and removal of cosmic rays. The wavelengths were calibrated through a dispersion solution from suitable lamp spectra. Flux calibration was obtained using standard stars observed at similar airmass on the same night. The spectra were further corrected for continuum atmospheric extinction and removal of telluric lines as far as possible. Keck spectra were reduced with the standard procedures described by \citet{Silverman2012}.\\

\section{Estimation of extinction and distance}
\label{sec:extinction}

\begin{figure}
	\includegraphics[width=\columnwidth]{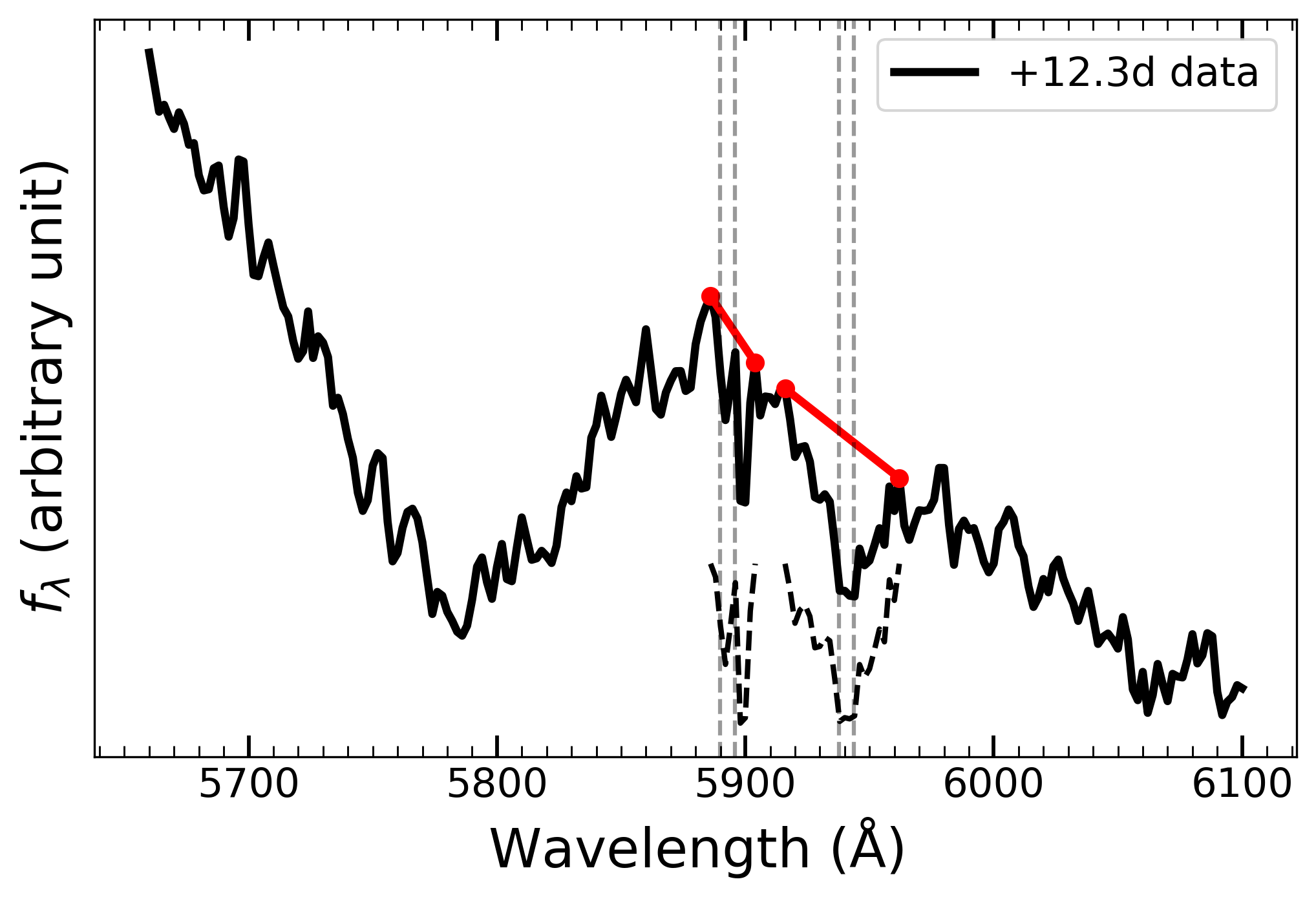}
    \caption{Identification of \ion{Na}{i~D} absorption lines from the host galaxy and the Milky Way. The spectrum shown was taken on 2018 Oct. 16 
(12.3\,d after explosion) with the 3\,m Shane telescope at Lick Observatory. The left vertical dashed lines denote \ion{Na}{i~D} absorption from the Milky Way and the right ones denote the trough due to the host galaxy. 
The red straight line is the presumed continuum. Bold dashed troughs under the spectrum are normalised absorption lines used to calculate equivalent width.}
    \label{fig:NaID_extinction}
\end{figure}

To estimate extinction both from the host galaxy and the Milky Way, we measured the equivalent width (EW) of \ion{Na}{i~D} from the Lick spectrum taken on 2018 Oct. 16. This high-quality spectrum is characterised by a blue continuum with shallow Balmer absorption lines. As shown in Figure~\ref{fig:NaID_extinction}, we can identify two \ion{Na}{i~D} systems in the 
spectrum; that from the Milky Way is at the rest wavelength of \ion{Na}{i~D}, and the redshifted one is caused by the host galaxy. Measurement of these two absorption lines gives us the EW of \ion{Na}{i~D} from the Milky Way as EW$_{\rm MW}=0.28$\,\AA\ and that from the host as EW$_{\rm host}=0.88$\AA. The results derived from four empirical formulae transforming the EW of \ion{Na}{i~D} to $E(B-V)$ are listed in Table~\ref{tab:reddening}. The median values and standard deviations are $E(B-V)_{\rm MW}=0.06\pm0.03$\,mag and $E(B-V)_{\rm host}=0.26\pm0.10$\,mag.\\


Meanwhile, we retrieve the Milky Way dust map from \citet{Schlegel1998} and find the line-of-sight reddening toward SN 2018hfm to be $E(B-V)=0.05$\,mag, consistent with the above result estimated from \ion{Na}{i~D} absorption lines. To double check the reddening due to the host galaxy, we adopt the Balmer-decrement method \citep{Osterbrock1989}. The principle of this method is that quantum physics determines the intrinsic flux 
ratio of $\rm H\alpha$ to $\rm H\beta$, and any deviation can be attributed to dust extinction. From the SDSS spectrum of the host galaxy \citep{Adelman-McCarthy2008}, as shown in Figure~\ref{fig:hostspec}, we measure a 
Balmer decrement of 4.56. As the SDSS spectrum was taken toward the host-galaxy centre and SN 2018hfm is quite near the centre, we do not expect a 
large deviation from the true value. According to Equation 4 of \citet{Dominguez2013}, we estimate the host-galaxy reddening to be $E(B-V)_{\rm host}=0.40$\,mag. This result is also consistnent with that estimated from the \ion{Na}{i~D} absorption. \\

Finally, we choose $E(B-V)$ of the Milky Way and the host galaxy to be 0.05\,mag and 0.26\,mag, respectively, resulting in a total reddening of $E(B-V)=0.31$\,mag. Assuming a total-to-selective extinction ratio $R_V=3.1$ \citep{Cardelli1989}, we obtain $A_V=R_V \times E(B-V)=0.96$\,mag and $A_B=1.27$\,mag\footnote{see Table 1 of\\ \url{http://www.astro.sunysb.edu/metchev/PHY517\_AST443/extinction\_lab.pdf}}, as well as $A_g=1.02$\,mag, $A_r=0.72$\,mag, and $A_i=0.53$\,mag \citep[Table 2]{Yuan2013}.\\

\textit{HyperLeda}\footnote{\url{http://leda.univ-lyon1.fr/ledacat.cgi?o=pgc1297331}}, a galaxy database, provides distance information for the host galaxy, PGC 1297331. Three distance moduli exist in the database: one is ``mod0'' calculated from their distance catalogue independent of redshift; another is ``modz'' computed from the redshift with H$_0=70$\,km\,s$^{-1}$\,Mpc$^{-1}$, $\Omega_M=0.27$, and $\Omega_{\Lambda}=0.73$; and the last is ``modbest,'' a weighted average of ``mod0'' and ``modz.'' We adopt ``modbest'' as our final distance modulus, giving a value of $32.70\pm 0.64$\,mag and corresponding to $D_L = 34.67\pm9.64$\,Mpc.

\begin{table*}
	\centering
	\caption{Reddening derived from different methods}
	\label{tab:reddening}
	\begin{threeparttable}
	\begin{tabular}{cccc} 
		\hline
		Relation    & $E(B-V)_{\rm MW}$\tnote{*} & $E(B-V)_{\rm host}$\tnote{**} & reference\\
		\hline
		0.25EW      & 0.07          & 0.22           & \citet{Barbon1990}\\
		0.16EW-0.01 & 0.03          & 0.13           & \citet{Turatto2003}\\
		0.51EW-0.04 & 0.10          & 0.41           & \citet{Turatto2003}\\
		0.43EW-0.08\tnote{***} & 0.04    & 0.30      & \citet{Poznanski2012}\\
		\hline
		median      & 0.06          & 0.26           &                     \\
		\hline
		standard deviation & 0.03   & 0.10           &                     \\
		\hline
	\end{tabular}
	
	  \begin{tablenotes}
	     \footnotesize
	     \item[*] EW of \ion{Na}{i~D} from the Milky Way is 0.28\,\AA.
	     \item[**] EW of \ion{Na}{i~D} from host is 0.88\,\AA.
	     \item[***] This formula requires EW $< 1$\,\AA. 
	  \end{tablenotes}
	\end{threeparttable}
\end{table*}

\section{Host Galaxy: PGC 1297331}
\label{sec:host}
\begin{figure}
	\includegraphics[width=\columnwidth]{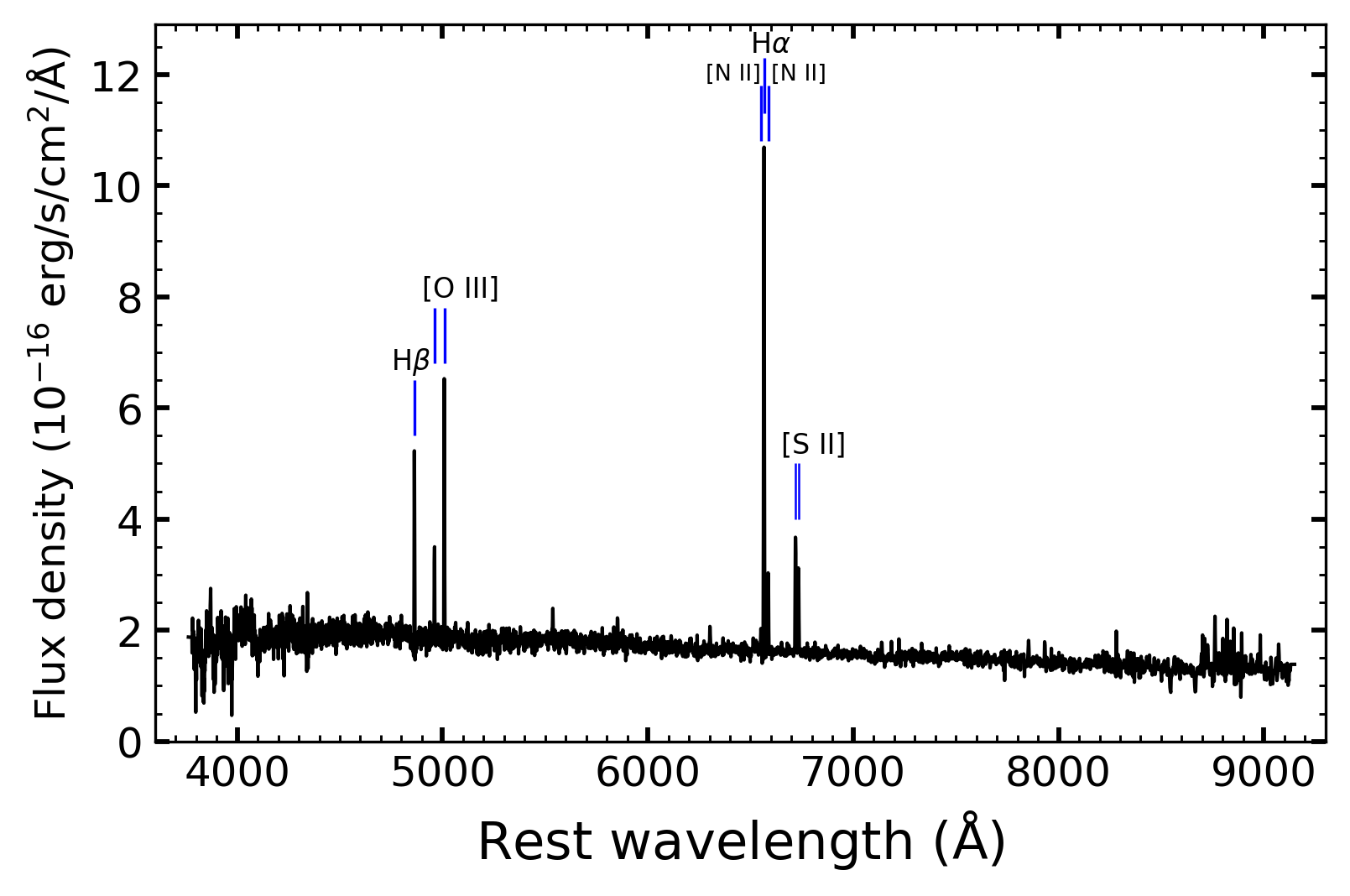}
    \caption{SDSS spectrum of the host galaxy of SN 2018hfm, PGC 1297331. 
Some strong lines are denoted.}
    \label{fig:hostspec}
\end{figure}

The host of SN 2018hfm, PGC 1297331, is classified as a transition-type dwarf (TTD) galaxy (\citealt{Koleva2013}; SDSS J0936 in their sample), which presents characteristics between those of irregular and elliptical galaxies. SDSS provides a spectrum taken toward the centre of PGC 1297331 \citep{Adelman-McCarthy2008}, as shown in Figure~\ref{fig:hostspec}.

\subsection{Metallicity}
\label{subsec:metallicity}
To determine the metallicity of the host environment, we measured the flux ratios of some strong emission lines in the SDSS spectrum of PGC 1297331, including $N2=\log_{10}([\ion{N}{ii}]\lambda 6584/\text{H}\alpha)$, $O3=\log_{10}([\ion{O}{iii}]\lambda 5007/\text{H}\beta)$, and $O3N2=\log_{10}(([\ion{O}{iii}]\lambda 5007/\text{H}\beta)/([\ion{N}{ii}]\lambda 6584/\text{H}\alpha))$. The results are $N2=-0.783$, $O3=0.286$, and $O3N2=1.069$. The values of $N2$ and $O3$ satisfy $O3<0.61/((N2-0.05)+1.3)$, 
suggesting no contamination from an active galactic nucleus \citep{Kauffmann2003}. We then calculated the gas-phase oxygen abundance as 8.388 dex (0.5\,Z$_\odot$) from $O3N2$ based on the relation derived by \citet{Pettini2004}. This value indicates that SN 2018hfm has a relatively metal-poor host environment compared to other SNe~II \citep{Dessart2014, Anderson2016}.\\

\subsection{Star-formation rate}
We retrieved a background-subtracted and flux-normalised far-ultraviolet (FUV) intensity map of PGC 1297331 from the \textit{GALEX} Catalog\footnote{\url{http://galex.stsci.edu/GR6/?page=mastform}}. We performed photometry on the galaxy with an elliptical aperture using the \texttt{photutils} Python package \citep{Bradley2017}, where the aperture parameters are 
adopted from \textit{HyperLeda}. The measured flux was converted into AB magnitude \citep{Oke_Gunn1983} using the zero point defined by Equation 3 
of \citet{Morrissey2007}. After applying Galactic and intrinsic extinction using Equation 4 of \citet{Karachentsev2013}, we obtained the extinction-corrected FUV magnitude of PGC 1297331, $m_{\rm FUV}^{\rm c} = 18.57$\,mag. We then transformed the FUV magnitude into the star-formation rate 
(SFR) using Equation 3 of \citet{Karachentsev2013}, namely $\log(\rm SFR[M_{\odot}\, yr^{-1}])=2.78 - 0.4m_{\rm FUV}^{\rm c}+2\log(D[\rm Mpc])$. 
Given $D=34.67$\,Mpc, the SFR can be calculated as $0.0270\, \rm M_{\odot} \cdot\,yr^{-1}$. \citet{Chang2015} report $\sim 0.021\,\rm M_{\odot}\,yr^{-1}$ as a median SFR estimate for the whole galaxy, thus verifying our result. \\

We adopt $0.0270\, \rm M_{\odot}\,yr^{-1}$ as the SFR of PGC 1297331. This SFR corresponds to a low CCSN rate of $\sim 1$ SN per 5000\,yr, which is 1--2 orders of magnitude below that of the hosts of the CCSNe discussed 
by \citet{Botticella2012}. 

\section{Photometric evolution}
\label{sec:phot}
The overall evolution of the SN 2018hfm light curves is shown in Figure~\ref{fig:LC}. These multiband light curves reveal four evolutionary stages: a rising phase, a plateau-like phase, a rapid-dropping transition, and a tail phase. For the plateau-like phase, SN 2018hfm presents special features  --- a fast decline rate and short duration. 

\subsection{Explosion date and rise time}
\label{subsec:expl}
\citet{Gutierrez2017a} describe two methods to determine the explosion epoch. One is to set it as the midpoint between the last nondetection date (MJD$_{\rm nondet}$) and the discovery date (MJD$_{\rm disc}$), along with the representative uncertainty determined by $(\rm MJD_{disc}-\rm MJD_{nondet})/2$. Another is  to perform a comparison between the observed spectra at early times and \textsc{SNID} templates \citep{2011ascl.soft07001B} to find a best-fit spectral phase. Applying these two methods to their 
SN~II sample, \citet{Gutierrez2017a} find a mean offset of 0.5\,d between 
them. These two methods have also been used by \citet{A14} to determine explosion epoch when they analysed their SN~II sample; However, they find an offset of 1.5\,d between the two methods. \\

The earliest detected photometric point of SN 2018hfm, obtained from ASASSN \texttt{Sky Patrol}, was taken on MJD = 58400.6 with a value of 14.97\,mag in the $V$ band. Given that the magnitude is brighter than any later ones in $V$, we consider this point to be the peak of the $V$ light curve. Eight days before this peak (i.e., MJD = 58392.6), an upper limit in $V$ ($>17.97$ mag) was also provided by ASASSN. After correcting for extinction, we converted the peak magnitude and the upper limit to an effective flux density $f$ of 9.0 and 0.57 (in units of $10^{-15}$\,erg\,s$^{-1}$\,cm$^{-2}$\,\AA$^{-1}$), respectively. We then fit a simple formula, 
$f(t) \propto t^2$ \citep{Arnett1982}, to the above two data points, and let $f(t) = 0$ to obtain the earliest possible explosion epoch, MJD = 
58389.9. Assuming this date as the MJD$_{\rm nondet}$ and the time of peak brightness as the MJD$_{\rm disc}$, we determine the explosion epoch as 
MJD $= 58395.2 \pm 5.3$, according to the first method.\\

\begin{table}
	\centering
	\caption{Best-fit spectral templates from \textsc{SNID}}
	\label{tab:SNID_epoch}
	\begin{threeparttable}
	\begin{tabular}{cccc} 
		\hline
		SN name & days from explosion & SN name & days from explosion\\
		\hline
		SN 2006bp & +12.59 & SN 1999em & +8.3\\
		SN 2006iw & +11.0 & SN 2008in & +4.0\\
		SN 2009bz & +9.0 & SN 1999gi & +7.7\\
		SN 2004fc & +9.0 & SN 2004et & +13.18\\
		\hline
		mean      & +9.4\tnote{*} &  $\sigma$ & 2.8\\
		\hline
	\end{tabular}
	   \begin{tablenotes}
	      \footnotesize
	      \item[*] The compared spectrum of SN 2018hfm was taken on MJD = 58407.5. Therefore, a phase estimate of $+9.4 \pm 2.8$\,d after explosion 
corresponds to an explosion epoch of MJD $= 58398.1 \pm 2.8$.
	   \end{tablenotes}
	\end{threeparttable}
\end{table}

In the second method, we compare our spectrum taken on MJD = 58407.5 with \textsc{SNID} templates. We focus on the wavelength between 3500\,\AA\ 
and 6000\,\AA, because spectral lines in this region evolved with time consistently, while the $\rm H\alpha$ profile at redder wavelengths varies between SNe so it does not aid in spectral matching \citep{Gutierrez2017a}. The best-fit template spectra are listed in Table~\ref{tab:SNID_epoch}. The mean value and standard deviation of the phase from these matched 
spectra thus gives us an alternative estimate of the explosion epoch, MJD 
$= 58398.1 \pm 2.8$. This value is offset by 2.9\,d from the first result, which is larger than the mean offset proposed by \citet{Gutierrez2017a} (0.5\,d) and \citet{A14} (1.5\,d). We attribute this large offset to a 
relatively low metallicity (0.5\,Z$_\odot$) of SN 2018hfm, as SNe with lower metallicity tend to exhibit metal lines of similar intensity at a phase later than their counterparts with higher metallicity \citep{Dessart2014}. \\
We adopt the more conservative (with the larger uncertainty) value, MJD $= 58395.2 \pm 5.3$, as the explosion epoch for SN 2018hfm.

\subsection{Light-curve parameters}
\label{subsec:LC_para}
Statistical studies of SNe~II are usually first based on parameters measured from $V$-band light curves (e.g., \citealt{A14,V16}). Thus, in this subsection, we perform a detailed analysis of the $V$ light curve of SN 2018hfm.\\

\begin{figure*}
	\includegraphics[width=2\columnwidth]{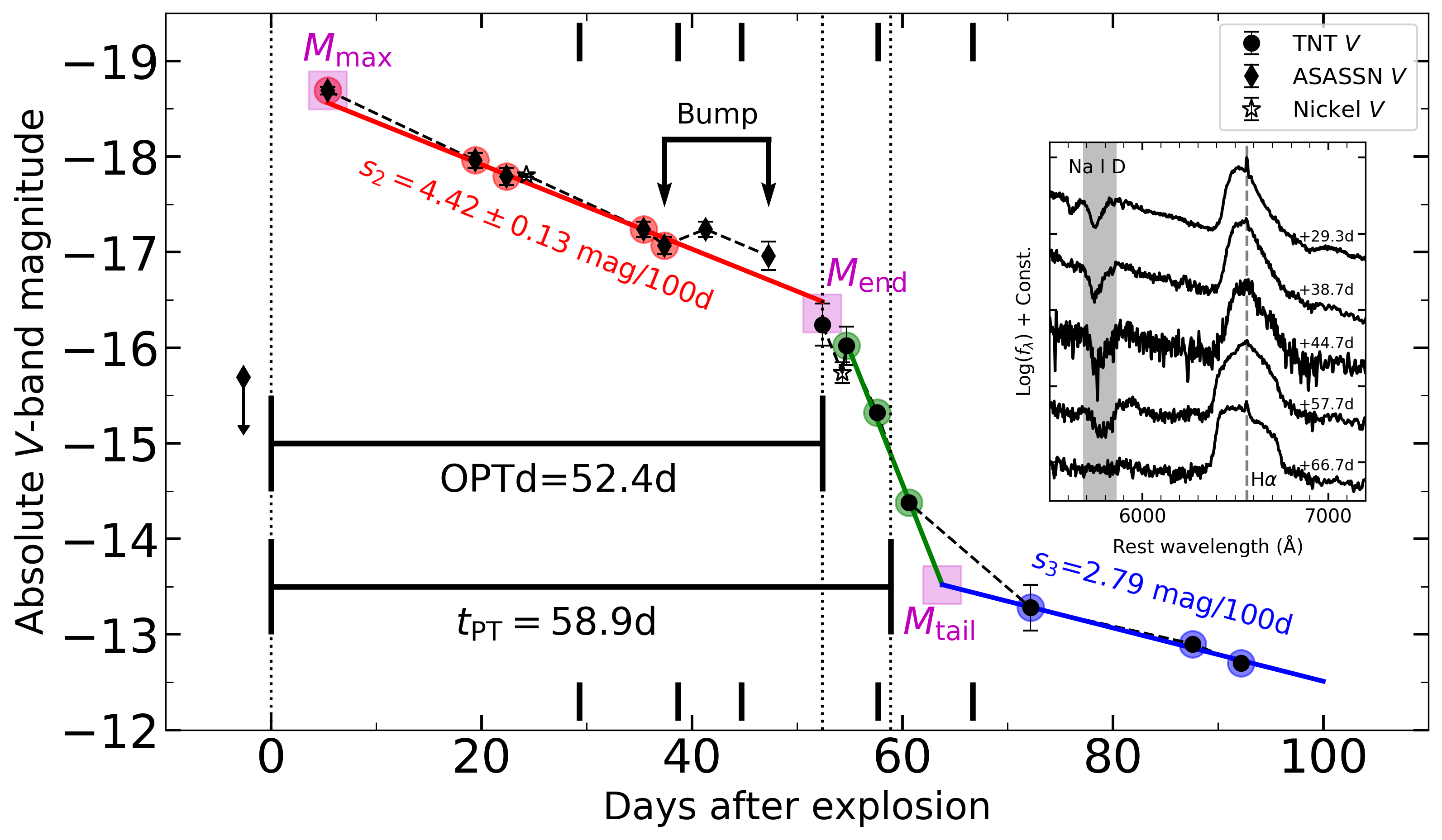}
    \caption{Measuring parameters from extinction-corrected absolute $V$-band light curve. Data are represented with black diamonds (from ASASSN), 
hollow stars (from Nickel), and circles (from TNT). Three straight lines (red/green/blue) are fitted to the data points circled with the same colour, to determine the decline rate of the plateau ($\rm s_2$) and of the tail ($\rm s_3$). Three magnitudes ($M_{\rm max}$, $M_{\rm end}$, and $M_{\rm tail}$) are denoted by magenta squares. Values of the optically-thick 
duration (OPTd) and $t_{\rm PT}$ are marked. The small bump at the end of the plateau-like phase is indicated by two linked arrows. Short black bars mark the spectral epochs shown in the inset.}
    \label{fig:analyse_Vband_LC}
\end{figure*}

Following the analysis method described by \citet{A14}, we select 5 points in the plateau-like phase, 3 points in the transition phase, and 3 
points in the tail phase, fitting each of them with a straight line as shown in Figure~\ref{fig:analyse_Vband_LC}. Note that the last two points at the plateau phase, which tend to form a small ``bump,'' are excluded in the fitting. The decline rates measured for the plateau-like phase and the tail phase are dubbed $s_2$ and $s_3$, respectively, according to the 
definition given by \citet{A14}. For SN 2018hfm, $s_2$ is measured to be $4.42\pm 0.13$\,mag\,(100\,d)$^{-1}$ and $s_3 = 2.79$\,mag\,(100\,d)$^{-1}$.\\

As mentioned in Section~\ref{subsec:expl}, we regard the first detected point as the peak of the $V$-band light curve, so that the maximum magnitude ($M_{\rm max}$) and the corresponding time ($t_{\rm max}$) can be determined. We assume that the ending of the plateau ($t_{\rm end}$) arrives +52.4\,d after explosion, thus the optically-thick duration (OPTd) is determined as shown in Figure~\ref{fig:analyse_Vband_LC}. To estimate the $V$-band magnitude at the end of the plateau ($M_{\rm end}$), we extrapolate the first straight line (the red one in Figure~\ref{fig:analyse_Vband_LC}) until $t_{\rm end}$ and get an inferred magnitude. We then average it with the observed magnitude at the same time (i.e., the first data point from TNT), and adopt the mean value as $M_{\rm end}$. For $t_{\rm tail}$ and $M_{\rm tail}$ (i.e., the beginning time and magnitude of the tail phase), we extrapolate the second (green) and the third (blue) lines in Figure~\ref{fig:analyse_Vband_LC}) to let them intersect each other, and then 
we take the corresponding values at the intersection point. The middle-plateau time ($t_{\rm mid}$) is calculated as $(t_{\rm max} + t_{\rm end})/2$ and the corresponding magnitude ($M_{\rm mid}$) is inferred from the best-fit straight line. All of the parameters mentioned above are listed in Table~\ref{tab:Vband_paramter}.\\

\begin{table}
	\centering
	\caption{Parameters measured from the absolute $V$-band light curve.}
	\label{tab:Vband_paramter}
	\begin{threeparttable}
	\begin{tabular}{cccccc} 
		\hline
		$t_{\rm max}$\tnote{*} & $M_{\rm max}$ & $t_{\rm end}$ & $M_{\rm end}$ & $t_{\rm tail}$ & $M_{\rm tail}$ \\
		\hline
		+5.4 & $-18.69 \pm 0.64$ & +52.4 & -16.36 & +63.8 & -13.52 \\
		\hline
		$t_{\rm mid}$ & $M_{\rm mid}$ & $s_2$\tnote{**} & $s_3$\tnote{**} & & \\
		\hline
		+28.9 & -17.52 & $4.42\pm 0.13$ & 2.79\tnote{$\Delta$} & & \\
		\hline
	\end{tabular}
	    \begin{tablenotes}
	       \footnotesize
	       \item[*] Days relative to explosion epoch (MJD = 58398.1).
	       \item[**] $s_2$ and $s_3$ are in units of mag\,(100\,d)$^{-1}$.
	       \item[$\Delta$]$s_3$ is obtained by fitting data points directly. 
We do not trust the data uncertainties in the tail phase.
	    \end{tablenotes}
	\end{threeparttable}
\end{table}

We caution that the determination of some of the above parameters is somewhat arbitrary, so we apply an alternative method to solve this problem. According to \citet{OlivaresE.2010} and \citet{V16}, light curves of 
SNe~II during the transition and the tail phase can be described well by the formula
\begin{equation}
    y(t)=\left(\frac{-a_0}{1+e^{(t-t_{\rm PT})/\omega_0}}\right) + (p_0\, t + m_0),
    \label{eq:tran_tail_fml}
\end{equation}
where the first term is a Fermi-Dirac function describing the shape of the transition and the second term is a straight line describing the evolution of the tail. Here, $a_0$ reflects the depth of the transitional drop, $t_{\rm PT}$ reflects the length of the plateau (which is 58.9\,d for 
SN 2018hfm; see also Figure~\ref{fig:analyse_Vband_LC}), $\omega_0$ represents the timescale of the transition phase, $p_0$ (in units of mag\,d$^{-1}$) has a meaning similar to that of $s_3$ (i.e., decline rate of the tail), and $m_0$ is an intercept related to the brightness of the tail phase. We fit all of the $BVgri$-band light curves during the transition and 
tail phase with the above formula; the results are listed in Table~\ref{tab:mutiband_para}.\\

\begin{table}
	\centering
	\caption{Fitting parameters (Eq. 1) of multiband light curves.}
	\label{tab:mutiband_para}
	\begin{tabular}{cccccc} 
		\hline
		 band & $a_0$ & $t_{\rm PT}$ & $\omega_0$ & $p_0$     & $m_0$\\
		      & (mag)       & (d)           & (d)           & (mag\,d$^{-1}$)   
& (mag)  \\
		\hline
		 $B$    & 2.00      & 57.4         & 1.74           & 0.0137    & -14.52\\
		 $V$    & 2.52      & 58.9         & 2.06           & 0.0278    & -15.30\\
		 $g$    & 1.94      & 57.6         & 1.50           & 0.0308    & -15.84\\
		 $r$    & 1.21      & 57.7         & 0.95           & 0.0294    & -17.07\\
		 $i$    & 2.16      & 58.5         & 2.08           & 0.0248    & -15.65\\
		\hline
	\end{tabular}
\end{table}

\begin{figure}
	\includegraphics[width=\columnwidth]{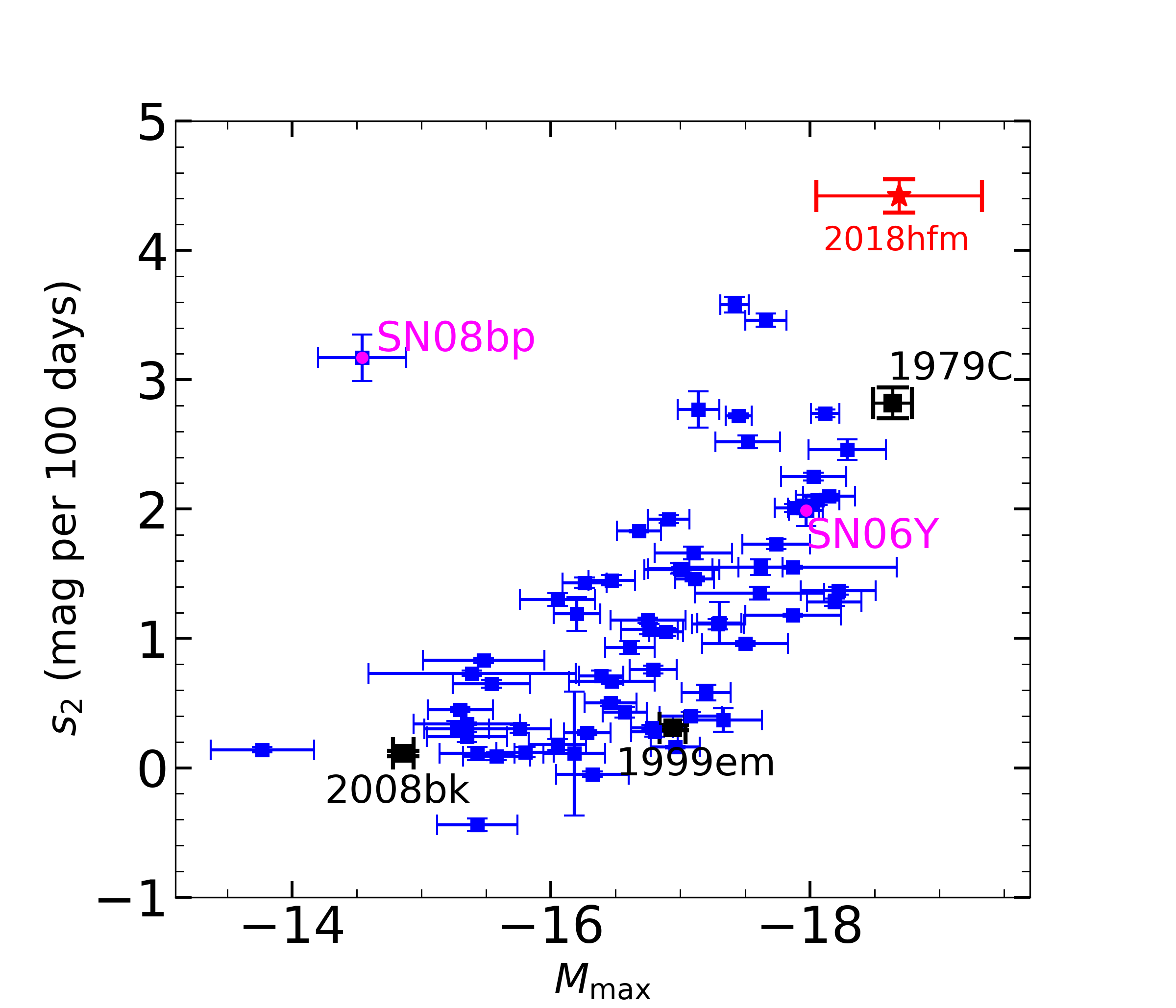}
    \caption{Relationship between absolute $V$-band magnitude at maximum brightness and decline rate of plateau-like phase inherited from \citet{A14}. Their 68 samples are denoted by blue squares with error bars. The subluminous SN 2008bk, prototype SN~IIP 1999em, and prototype SN~IIL 1979C are denoted by black squares. Fuchsia dots label the outlier SN 2008bp and the peculiar SN 2006Y. SN 2018hfm is denoted by a red star.}
    \label{fig:Vmax_s2}
\end{figure}

Normal SNe~IIP usually have a plateau phase lasting for $\sim 100$\,d. The estimate of OPTd (52.4\,d) and $t_{\rm PT}$ (58.9\,d) for SN 2018hfm, however, indicate that it belongs to short-plateau SNe \citep{Hiramatsu2020}. SNe~II with a short plateau tend to have large decline rates during the plateau-like phase (see Figure~5(a) of \citealt{V16}). This is the case for SN 2018hfm, which indeed has a large value of $s_2$ ($4.42 \pm 0.13$\,mag\,(100\,d)$^{-1}$). A more significant and widely accepted relation is between $M_{\rm max}$ and $s_2$, where more-luminous SNe~II tend to decline faster \citep{A14}. To show that SN 2018hfm conforms to this 
relation, we replot Figure 7 of \citet{A14} by including SN 2018hfm (see Figure~\ref{fig:Vmax_s2}). The upper-right location of SN 2018hfm reveals its exceptionally high luminosity and rapid decline rate. More cases like SN 2018hfm will extend the data coverage and further examine the relation. \\

The decline rates during the tail phase of SN 2018hfm, whether in $V$ (reflected by $s_3$) or in other bands (reflected by $p_0$ values listed in Table~\ref{tab:mutiband_para}), are all larger than the expected value of 
the $\rm ^{56}Co$-to-$\rm ^{56}Fe$ decay rate if gamma-ray photons are effectively trapped (i.e., 0.98\,mag\,(100\,d)$^{-1}$). This indicates that 
leakage of gamma-ray photons occurs, or that the tail-phase light curves are influenced by other energy source (e.g., CSI).\\

One may notice that a small ``bump'' emerges at the end of the plateau (see Figure~\ref{fig:analyse_Vband_LC}). It is observed not only in $V$ but also in $g$ (see Figure~\ref{fig:LC}). To explore the origin of such a bump, which is not normal in SNe~II, we select five spectra obtained around the time of the bump ---  one before (+29.3\,d), two during (+38.7\,d, +44.7\,d), one after (+57.7\,d) but during the transition phase, and one (+66.7\,d) in the tail phase. These spectra are shown in the inset of Figure~\ref{fig:analyse_Vband_LC}. \\

Among the five spectra, the one at +29.3\,d has a blackbody-like continuum and P~Cygni profiles of hydrogen and metal lines, similar to other SNe~II in their photospheric phase. In the +38.7\,d and +44.7\,d spectra, the 
top of the $\rm H\alpha$ emission becomes not so smoothly round. At +57.7\,d, when the luminosity decreases but does not yet enter the tail phase, 
the spectrum exhibits a very broad and bell-shaped profile of $\rm H\alpha$ emission, with metal lines still existing. In the +66.7\,d spectrum, metal lines such as \ion{Na}{i~D} are absent, and the continuum becomes flat and featureless; only $\rm H\alpha$ is present, with a box-like shape and a prominent flux deficiency on the red side.\\

The photometric and spectroscopic features described above can be explained by the coexistence of two main energy sources. One is the thermal energy deposited in the SN envelope by the explosion shock, which corresponds to the main energy source powering the plateau phase for those normal SNe~II without CSI. The other energy source comes from the interaction between the SN ejecta and the CSM. For convenience, we call the former as thermal energy component and the latter as CSI component. Note that when the CSI occurred is a problem worth to be discussed. For SN 2018hfm, the thermal energy component dominates during the plateau phase, so spectra during this phase show similar features to other normal SNe~II in the photospheric phase, i.e., P-Cygni profiles. The small bump emerging at the end of the plateau can be explained by that the CSI at this time has occurred, thus extra energy from interaction is superimposed on the thermal energy; however, the CSI energy component is still weak at this time, so we do not see obvious box-like profile of $\rm H\alpha$. In the transition phase, the thermal energy component becomes relatively weak, and the interaction component is increasingly important, we thus see a transitional bell-shaped $\rm H\alpha$ emission line in the +57.7\,d spectrum. At the tail phase, when the thermal energy is exhausted and the CSI component becomes dominant, the spectra exhibit box-like $\rm H\alpha$ emission lines with a flat, featureless continuum. \\

To get an idea about the contribution of CSI, we investigate the fraction of H$\alpha$ flux due to CSI at five epochs ($t=$+29.3d, +38.7d, +44.7d, +57.7d and +66.7d). We first calibrate these five spectra  with the corresponding photometric data, and then we estimate the H$\alpha$ flux by integrating the spectra between 6267\,\AA~and 6913\,\AA~. This wavelength range corresponds to the inner and outer layers of CSI emission region (see Appendix \ref{sec:preprossing}). The H$\alpha$ flux measured at $t=$+66.7d is $\sim 2.15\times 10^{-13}$\,erg\,s$^{-1}$\,cm$^{-2}$, which should be resulted primarily from the CSI. Assuming that the H$\alpha$ flux due to CSI does not change significantly during the period from $t=$+29.3d to $t=$+66.7d, then we can roughly estimate the H$\alpha$-flux fraction due to CSI at other four epochs. The results are 12.6\% at $t=$+29.3d, 17.4\% at $t=$+38.7d, 19.7\% at $t=$+44.7d, and 40.9\% at $t=$+57.7d, respectively. One can clearly see the CSI component plays an increasingly important role in shaping the H$\alpha$ line.

\subsection{Comparison with other SNe}
\label{subsec:compVLCs}
\begin{figure*}
	\includegraphics[width=2\columnwidth]{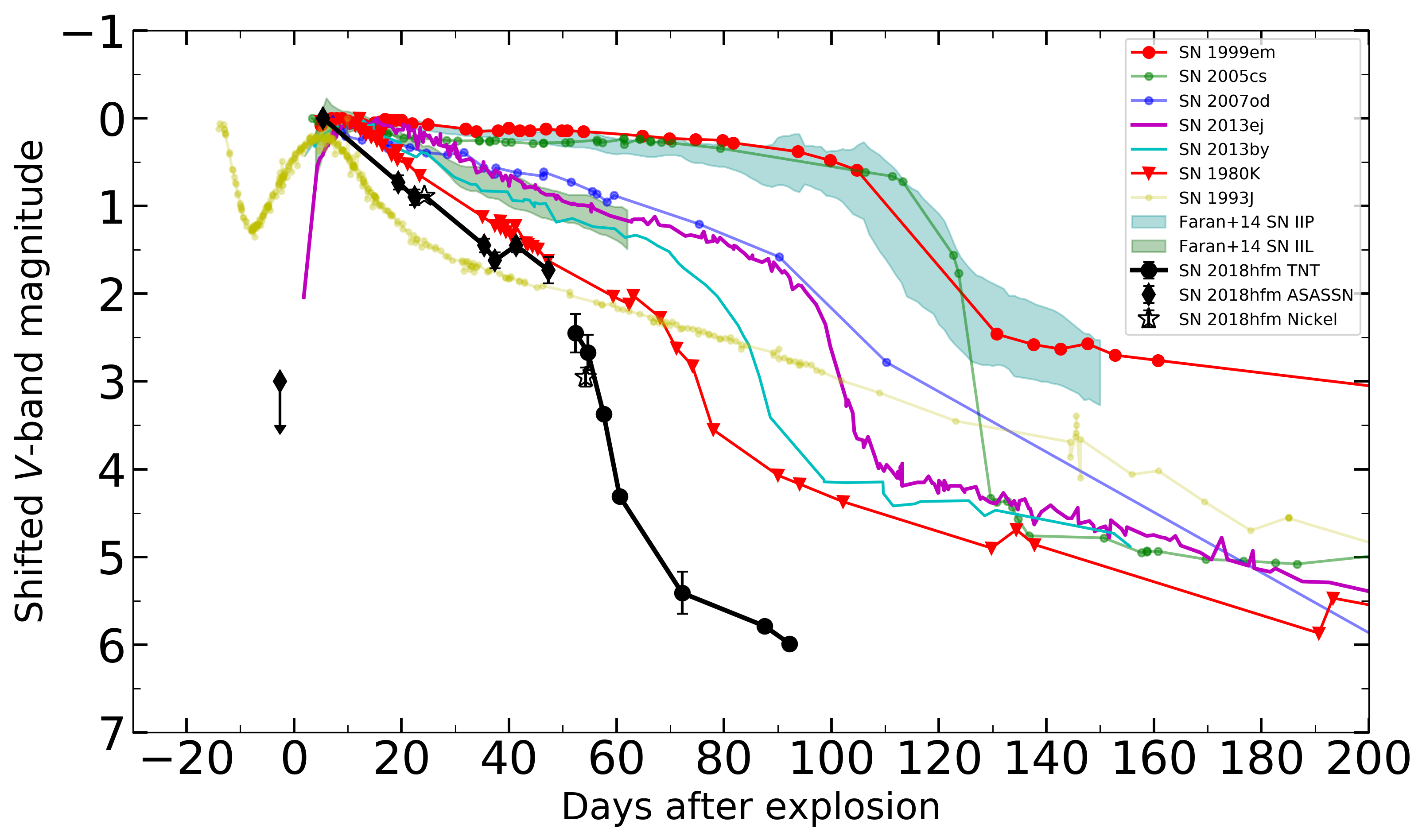}
    \caption{Extinction-corrected absolute $V$-band light curve of SN 2018hfm along with that of other SNe, including prototype SN~IIP 1999em \citep{Elmhamdi2003}, prototype SN~IIL 1980K \citep{Barbon1982, Buta1982, Marano1980}, prototype SN~IIb 1993J \citep{Richmond1994, Richmond1996}, SN 2005cs \citep{Pastorello2009}, SN 2007od \citep{Inserra2011}, SN 2013ej \citep{Huang2015}, SN 2013by \citep{Valenti2015}, and template light curves 
of SNe~II from \citet{Faran2014b}. All of the light curves have been shifted vertically to their $V$-band maximum. SN 1993J is also shifted horizontally so that its maximum epoch is aligned with that of SN 2018hfm. The light curves of some SNe are taken from \url{https://sne.space} \citep{Guillochon2017}.}
    \label{fig:compare_Vband}
\end{figure*}

In Section~\ref{subsec:LC_para} we measured the light-curve parameters and compared them with those of statistical samples of \citet{A14} and \citet{V16}. The comparison has revealed some photometric uniqueness of SN 2018hfm, such as a high luminosity at maximum brightness, a short plateau duration, and a large decline rate of the plateau-like phase. To further examine the peculiarities of this SN~II, in Figure~\ref{fig:compare_Vband} we compare the $V$-band light curve of SN 2018hfm with that of some other 
SNe~II. \\

As shown in Figure~\ref{fig:compare_Vband}, SN 1999em, as a prototype of SNe~IIP, shows a long plateau ($\sim 100$\,d) with a very slow decline during the plateau phase. SN 2005cs, similar to SN 1999em, also has a typical long plateau. However, SN 2013ej and SN 2013by show a faster decline rate during the plateau-like phase, being in the range of the SN~IIL template light curves. These two SNe were observed sufficiently well to catch the final rapid drop at the end of the plateau. SN 2007od tends to have an intermediate plateau drop between that of SNe~IIP and SNe~IIL. As a prototype of SNe~IIL, SN 1980K exhibits a plateau decline rate even larger than that of most SNe~IIL examined by \citet{Faran2014b}. SN 1993J is a prototype of SNe~IIb, which has the most rapid decline rate after its main peak in comparison with other SNe~II shown in Figure~\ref{fig:analyse_Vband_LC}. Regarding the post-peak evolution, one can see that SN 2018hfm lies between SN 1980K and SN 1993J. \\

As discussed above, light curves of all the SNe~II (except SN 1993J) are found to show a four-stage evolutionary sequence. This result is consistent with the fact that progressively more evidence suggests that historically-classified SNe~IIP and SNe~IIL actually belong to a continuous distribution. Their observational differences are related to the mass of the remaining hydrogen envelope and hence to the mass-loss history of the progenitor stars \citep{A14, V16, Gutierrez2017a}. Luminous SNe~II with short plateaus and large decline rate are likely linked to progenitors having a low-mass envelope, while those with a long plateau, slow decline, and low luminosity are related to progenitors having a high-mass envelope. On the other hand, SNe~IIb are believed to retain a very low-mass hydrogen envelope before core collapse \citep[][and references therein]{Kilpatrick2017}, which may explain the fact that in Figure \ref{fig:compare_Vband} the light curve of SN 1993J declines fastest after its main peak. The light curve of SN 2018hfm shows a large post-peak decline, which indicates that its progenitor tends to hold a low-mass hydrogen envelope before exploding.

\subsection{Bolometic light curve and explosion parameters}
\label{subsec:boloLC}
\begin{figure}
	\includegraphics[width=\columnwidth]{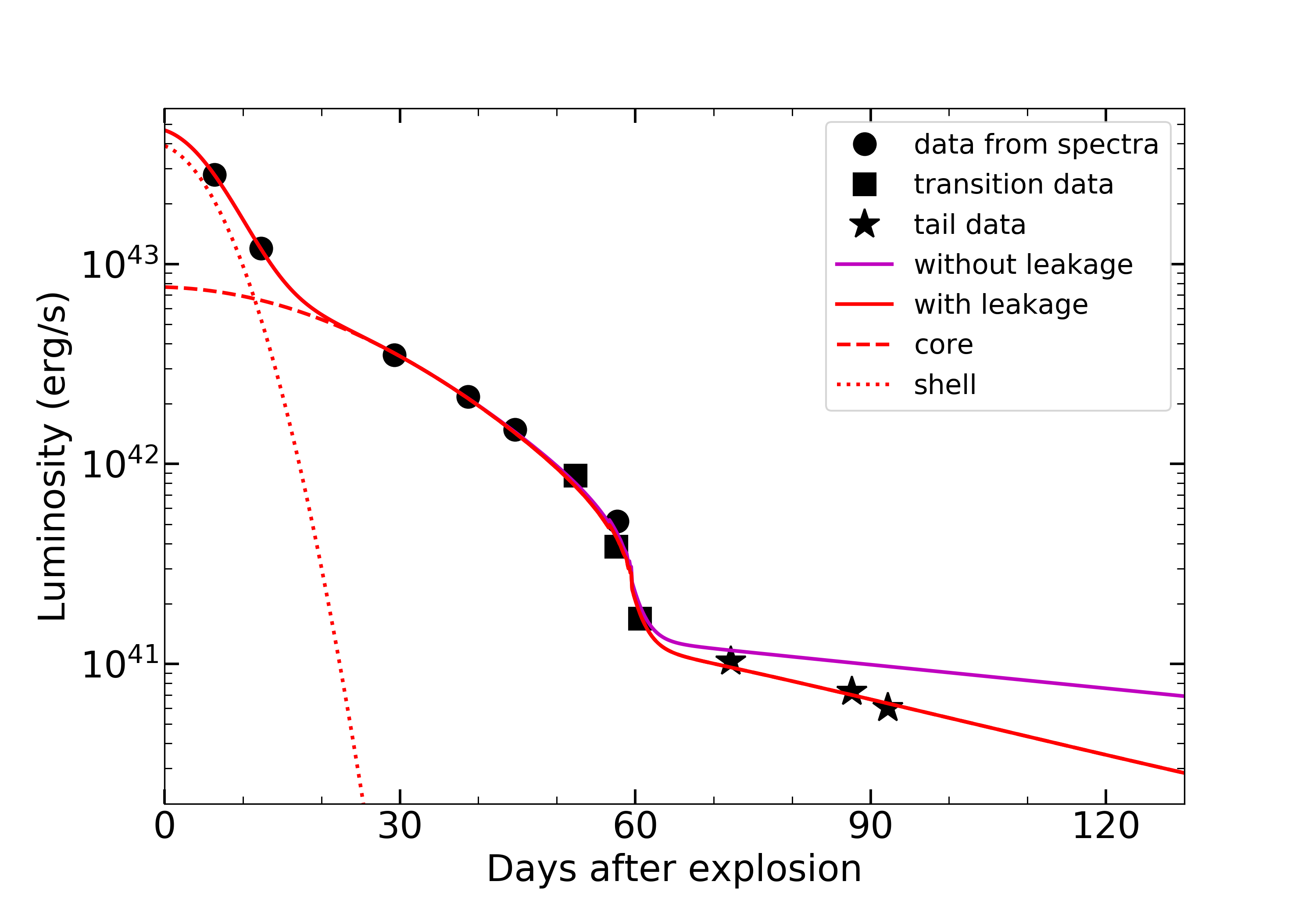}
    \caption{Bolometric light curve of SN 2018hfm (black markers) with LC2 models from \citet{Nagy2016} (red and magenta lines). Parameters related to these lines are listed in Table~\ref{tab:bolo_para}.}
    \label{fig:bolo_LC}
\end{figure}


In this subsection, we construct the bolometric light curve of SN 2018hfm 
(see Figure~\ref{fig:bolo_LC}) and discuss its properties. During the plateau-like phase we only have $V/g$-band data, but we have collected relatively good-quality spectra (i.e., +6.4\,d, +12.3\,d, +29.3\,d, +38.7\,d, and +44.7\,d). We calibrate\footnote{using a Python package \texttt{pysynphot} \\ \url{https://pysynphot.readthedocs.io/en/latest/spectrum.html\#renormalization}} these spectra by the $V$-band magnitude inferred from the best-fit straight line (the red line shown in Figure~\ref{fig:analyse_Vband_LC}). Note that the straight line does not describe the small bump seen 
in the ASASSN $V$ and $g$ bands, hence one would not expect a bump feature in the bolometric light curve.

After flux calibration, we perform a blackbody fit to the spectra and integrate it from 1216\,\AA\ to infinity; flux at wavelengths shorter than 1216\,\AA\ is omitted owing to absorption by the Lyman series \citep{Zhangjujia2020}. We do the same thing for the +57.5\,d spectrum, taken during the transition phase, but integrate it from 3600\,\AA\ to infinity because blanketing effects of metal lines play a significant role when 
the temperature decreases. Spectra taken during the tail phase are discarded because the SN was very faint and the continuum could be contaminated 
by the host galaxy. Multiband ($BVgi$) photometric data taken with the TNT during the transition phase were transformed to spectral energy distributions (SEDs) and then fitted with a blackbody. Similarly, the integration covers the wavelength range from 3600\AA\ to infinity. Note that the $r$-band data are excluded because they are severely influenced by 
the $\rm H\alpha$ emission produced by CSI. During the tail phase, the bolometric luminosity is reconstructed using the $V$-band data through the following equation \citep{Bersten2009, Zhangjujia2020}:
\begin{equation}
    \log_{10} L = -0.4[BC + V -A_{V} + 11.64] + \log_{10} (4\pi D^2),
	\label{eq:BH09}
\end{equation}
where $L$ is the luminosity in units of erg\,s$^{-1}$, $D$ is the distance to the SN in units of cm, and $BC$ is the bolometric correction ($-0.7$\,mag).\\

We also present in Figure~\ref{fig:bolo_LC} the bolometric light curves generated by the semi-analytical two-component model (LC2) of \citet{Nagy2016}. This model, assuming that the emission can be produced by an inner-core part and an outside-shell part, can be used to investigate explosion 
parameters of SNe~IIP/IIL or SNe~IIb. The LC2 parameters are listed in Table~\ref{tab:bolo_para}. It is not surprising that the model with gamma-ray leakage (i.e., $A_g \approx 9000$\,d$^2$) reproduces the tail if we recall that the values of $s_3$ and $p_0$ (see Tables~\ref{tab:Vband_paramter} and \ref{tab:mutiband_para}) are all larger than 0.98\,mag\,(100\,d)$^{-1}$.\\

\begin{table*}
	\centering
	\caption{LC2 model parameters of SN 2018hfm.}
	\label{tab:bolo_para}
	\begin{threeparttable}
	\begin{tabular}{cccccccccc} 
		\hline
		Parameters\tnote{*} & $R_0$ ($10^{13}$cm) & $M_{\rm ej}$ (M$_\odot$) & $M_{\rm Ni}$ (M$_\odot$) & $T_{\rm rec}$ (K) & $E_{\rm kin}$ ($10^{51}$\,erg) & $E_{\rm th}$ ($10^{51}$\,erg) & $\alpha$ & $\kappa$ (cm$^2$\,g$^{-1}$) & $A_{\rm g}$ (d$^2$) \\
		\hline
		core  & 10 & 1.17 & 0.015 & 6000 & 0.22 & 0.20 & 1.6 & 0.28 & 9000/$+\infty$\tnote{**} \\
		\hline
		shell & 20 & 0.11 & 0     & 0    & 0.1  & 0.1  & 0.0 & 0.34 & $+\infty$ 
\\
		\hline
	\end{tabular}
	    \begin{tablenotes}
	       \footnotesize
	       \item[*] $R_0$ is the initial radius of the ejecta,  $M_{\rm ej}$ 
is the ejected mass, $M_{\rm Ni}$ is the initial nickel mass, $T_{\rm rec}$ is the recombination temperature, $E_{\rm kin}$ is the initial kinetic 
energy,  $E_{\rm th}$ is the initial thermal energy, $\alpha$ is the density profile exponent, $\kappa$ is the opacity, and $A_{\rm g}$ is the gamma-ray leakage exponent.
	       \item[**]  $A_{\rm g} = +\infty$ is for the magenta line in Figure~\ref{fig:bolo_LC}, representing gamma-ray photons that are fully trapped, while $A_{\rm g}=9000$ is for the red line in Figure~\ref{fig:bolo_LC}, representing leakage of the gamma rays.
	    \end{tablenotes}
	\end{threeparttable}
\end{table*}

Among the LC2 parameters, the most important ones are $R_0$, $M_{\rm ej}$, and $E_{\rm tot}=E_{\rm kin}+E_{\rm th}$, which control the morphology of the plateau-phase light curve, and $M_{\rm Ni}$, which determines the tail phase. To examine these parameters we also estimate the explosion parameters ($E_{\rm exp}$, $M_{\rm ej}$, $R_{\rm pSN}$) using the following simple approximation formula \citep{Litvinova1985, Zhangtm2006}:
\begin{equation}
  \log_{10}
  \left(
  \begin{bmatrix}
      E_{\rm exp} \\
      M_{\rm ej}\\
      R_{\rm pSN} \\
  \end{bmatrix}
  \right)
  =
  \begin{bmatrix}
      0.135 & 2.34 & 3.13 \\
      0.234 & 2.91 & 1.96 \\
      -0.572 & -1.07 & -2.74 \\
  \end{bmatrix}
  \cdot
  \begin{bmatrix}
  M_V \\
  \log_{10}(\Delta t) \\
  \log_{10}(v_{\rm ph})\\
  \end{bmatrix}
  -
  \begin{bmatrix}
  4.205 \\
  1.829 \\
  3.350 \\
  \end{bmatrix}
  ,
  \label{eq:expl_para}
\end{equation} 
where $M_V$ (in units of mag) is the $V$-band absolute magnitude during the plateau phase, $\Delta t$ (days) is the length of the plateau, and $v_{\rm ph}$ ($10^3$\,km\,s$^{-1}$) is the photospheric velocity measured +50\,d after explosion, while $E_{\rm exp}$ ($10^{51}$\,erg), $M_{\rm ej}$ ($\rm M_{\odot}$), and $R_{\rm pSN}$ ($\rm R_\odot$) represent the explosion energy, ejecta mass, and initial radius of pre-supernova star, respectively. For SN 2018hfm, owing to a large slope of the plateau, we adopt the magnitude at the midpoint of the plateau (namely $M_{\rm mid}$ in Table~\ref{tab:Vband_paramter}) as $M_V$. Also, the OPTd marked in Figure~\ref{fig:analyse_Vband_LC} is regarded as the length of the plateau ($\Delta 
t$). The value $v_{\rm ph}$ on +50\,d is inferred from the velocity evolution of \ion{Fe}{ii} $\lambda$5169 measured from spectra (see Figure~\ref{fig:measure_spec}). The parameters calculated through Equation~\ref{eq:expl_para} are listed in Table~\ref{tab:expl_para}, which are consistent with those given by the LC2 model; both indicate a low ejecta mass and low explosion energy for SN 2018hfm compared with SN 1999em or SN 2018zd. (See 
Table~3 of \citealt{Zhangjujia2020}; for SN 1999em, they give $M_{\rm ej} 
= 13.50$\,M$_\odot$ and $E_{\rm tot}=1.88\times 10^{51}$\,erg in the core part; for SN 2018zd, they give $M_{\rm ej} = 9.80$\,M$_\odot$ and $E_{\rm tot}=4.10 \times 10^{51}$\,erg in the core part.)

\begin{table}
    \centering
    \caption{Explosion parameters calculated from Eq.~\ref{eq:expl_para}}
    \label{tab:expl_para}
    \begin{threeparttable}
      \begin{tabular}{ccc}
      \hline
      $M_V$ (mag) & $\Delta t$ (d) & $v_{\rm ph}$ ($10^3$\,km\,s$^{-1}$) \\
      \hline
      -17.52     & 52.4             & 3.8                \\
      \hline
      $E_{\rm exp}$ ($10^{51}$\,erg) & $M_{\rm ej}$ ($\rm M_\odot$) & $R_{\rm pSN}$ ($\rm R_\odot$)\\
      \hline 
       0.19      & 1.63            &  1750\tnote{*}          \\
      \hline
       \end{tabular}
       \begin{tablenotes}
          \footnotesize
          \item[*] 1750\,$\rm R_\odot$ = $12.18 \times 10^{13}$\,cm
       \end{tablenotes}
     \end{threeparttable}
\end{table}

\subsection{Evolution of colour curves}
\label{subsec:color_curve}
\begin{figure}
	\subfigure[]{
	    \includegraphics[width=\columnwidth]{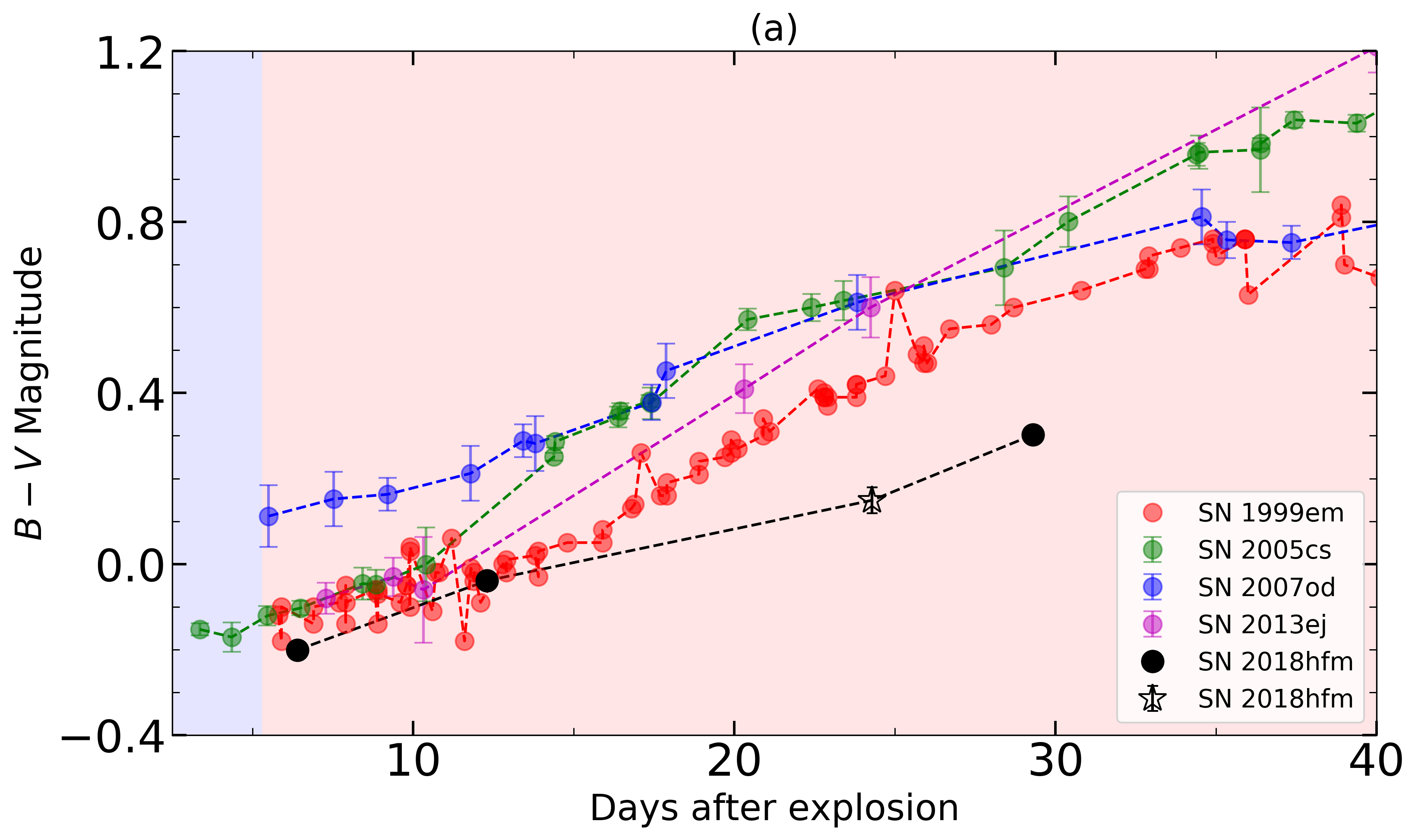}
	}
	\quad
	\subfigure[]{
	    \includegraphics[width=\columnwidth]{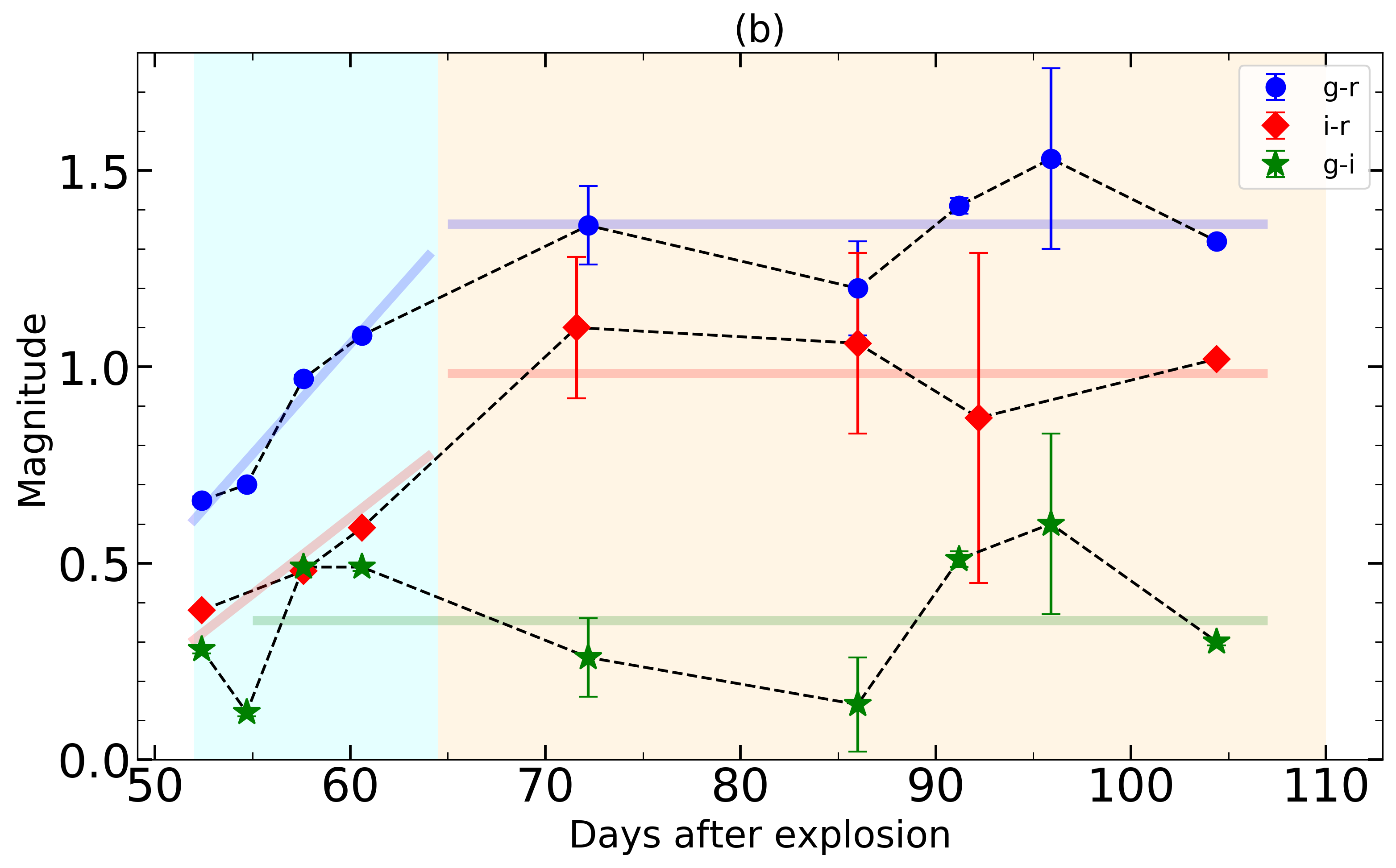}
	}
    \caption{Intrinsic colour curves of SN 2018hfm. Shaded areas represent four evolutionary stages (blue for rise, red for plateau-like phase, cyan for transition, and yellow for tail phase). (a) $B-V$ colour of SN 2018hfm calculated from Nickel photometric data (the hollow star) and from synthetic photometry performed on the early-time spectra (black filled circles), together with $B-V$ colour of SN 1999em \citep{Elmhamdi2003}, SN 2005cs \citep{Pastorello2009}, SN 2007od \citep{Inserra2011}, and SN 2013ej \citep{Huang2015}. (b) $g-r$, $i-r$, and $g-i$ colour curves of SN 2018hfm at the transition and tail phases using photometric data from the TNT. Straight lines show the evolutionary trend of the colour curves.}
    \label{fig:color_curve}
\end{figure}

SN 2018hfm was not well sampled photometrically within one month after the explosion; only the Lick/Nickel telescope took a night of data. However, we collected relatively good-quality spectra during this period. This allows us to perform synthetic photometry with these spectra and obtain the $B-V$ colour. The result is plotted in Figure~\ref{fig:color_curve}(a), 
together with those of the comparison samples, including SN 1999em, SN 2005cs, SN 2007od, and SN 2013ej. As  can be seen, SN 2018hfm is bluer than 
the comparison SNe, suggesting a higher temperature in the early phase. \\

The colour evolution $g-r$, $i-r$, and $g-i$ during the transition and tail phases are shown in Figure~\ref{fig:color_curve}(b). The value of $g-i$ stays almost unchanged during this time, consistent with the fact that the temperature stays stable and the continuum of the spectra appears very flat. By contrast, values of $g-r$ and $i-r$ both increase (evolving redward) during the transition phase, due to the internal energy deposited by the explosion shock falling and the CSI component gradually dominating the emission. The spectrum from CSI is characterised by strong $\rm H\alpha$ emission, so the $r$ band is brighter than the other bands. During the tail phase, when the CSI component is exposed completely, $g-r$ and $i-r$ stay nearly constant. This means the $\rm H\alpha$ emission and the continuum come from the same energy source (i.e., CSI), which further indicates that the mass of $\rm ^{56}Ni$ is likely very small and has no significant influence on the tail-phase light curves.

\section{Spectroscopic evolution}
\label{sec:spec}
The complete spectral evolution of SN 2018hfm is displayed in Figure~\ref{fig:spectra}, spanning from +6.4\,d to +389.4\,d after explosion. The first spectrum was taken on +6.4\,d, only one day after $V$-band maximum; it is characterised by a very blue and featureless continuum. Applying a blackbody fit to this spectrum indicates a high temperature of $T_{\rm bb} 
= 18,628 \pm 94$\,K. The second spectrum, taken +12.3\,d after explosion, also has a blue continuum but with the appearance of \ion{He}{i} $\lambda$5876 absorption, which has a velocity of $\sim 7000$\,km\,s$^{-1}$. By $\sim 1$\,month after explosion, when the plateau-like phase arrives at 
its midpoint, metal lines emerge in the spectra. During the transition phase of the light curve, the $\rm H\alpha$ emission shows a strange bell-shaped profile (see discussion in Sec.\ref{subsec:LC_para}). When the SN evolves into the tail phase, metal lines become invisible, while broad, boxy $\rm H\alpha$ emission and relatively faint [\ion{Ca}{ii}] $\lambda\lambda$7291, 7323 emission dominate the spectra.

\subsection{Comparison with other SNe II}
\label{subsec:compspec}
\begin{figure*}
	\includegraphics[width=2\columnwidth]{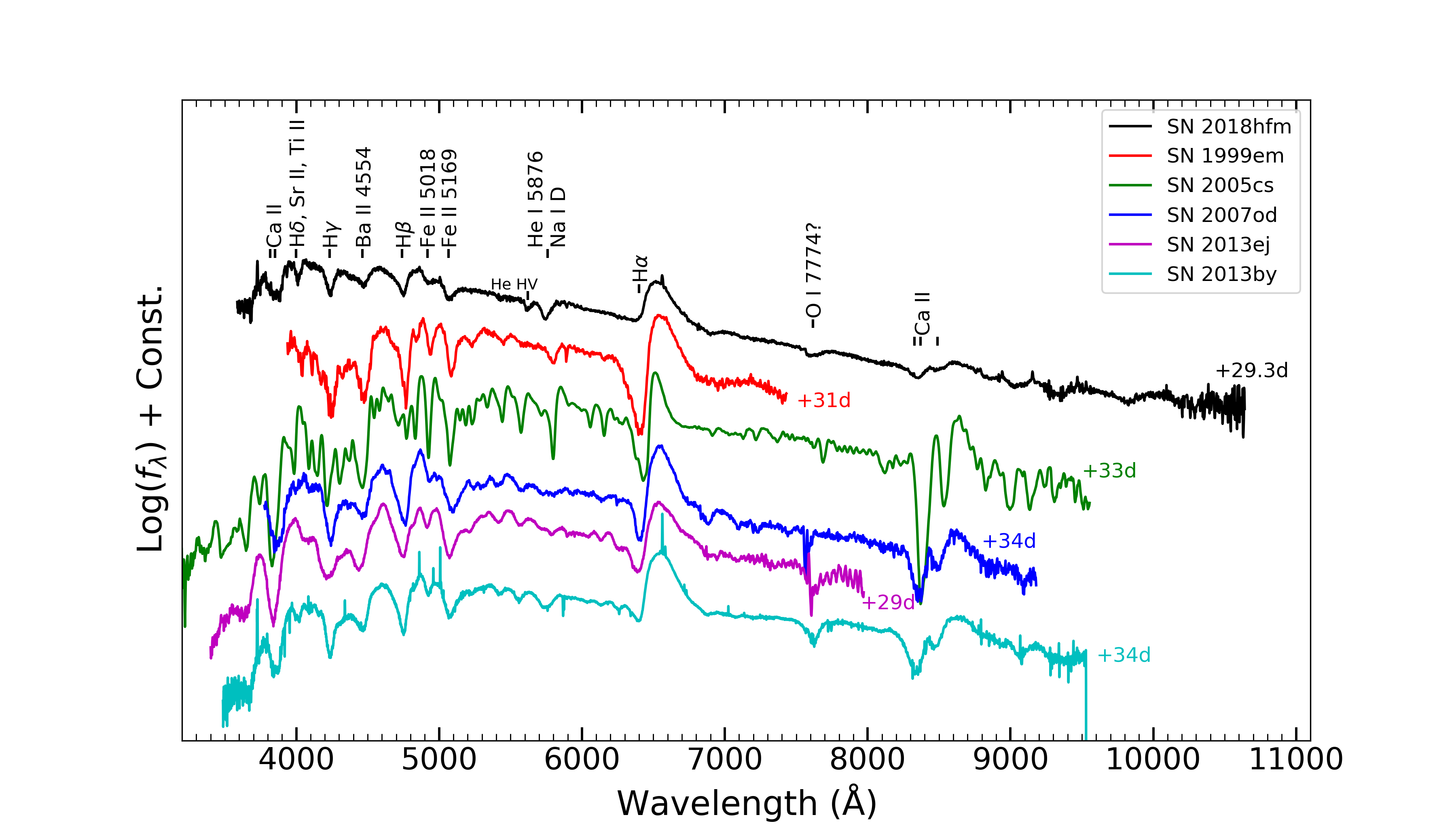}
    \caption{Photospheric spectrum of SN 2018hfm (taken at +29.3\,d) compared with that of other SNe~II at a similar epoch, including SN 1999em \citep{Elmhamdi2003}, SN 2005cs \citep{Pastorello2006}, SN 2007od \citep{Inserra2011}, SN 2013ej \citep{Huang2015}, and SN 2013by \citep{Valenti2015}. The epoch of each spectrum is denoted on the right side with the same colour. Line identifications are marked by short bars. A high-velocity feature of \ion{He}{i} is also marked.}
    \label{fig:compspec1}
\end{figure*}

In Figure~\ref{fig:compspec1}, we compare the +29.3\,d spectrum of SN 2018hfm with spectra of SN 1999em, SN 2005cs, SN 2007od, SN 2013ej, and SN 2013by at similar phases. With the comparison, we identify the Balmer series and \ion{He}{I} absorption in the spectrum, along with metal lines such as \ion{Fe}{ii}, \ion{Ca}{ii}, \ion{Ba}{ii}, and \ion{O}{i}. We notice that a small notch exists on the left side of \ion{He}{i} $\lambda$5876 in the +29.3\,d spectrum. We checked all of the spectra of SN 2018hfm, and find that this notch also appeared in the +12.6\,d spectrum, but it disappeared in the spectral series by +29.3\,d and thereafter. This change coincides with a decrement in the temperature, and the notch is likely a high-velocity feature of helium. Moreover, we notice that SN 2018hfm has shallower and fewer metal absorption lines than other SNe~II, consistent with the low-metallicity environment \citep{Dessart2014}. The absorption component of the P~Cygni $\rm H\alpha$ profile tends to be not well developed for SNe~II whose light curve has a fast post-peak decline rate \citep{Schlegel1996, Faran2014b}, as indicated by the spectra of SN 2013ej, SN 2013by, and SN 2018hfm.\\

\begin{figure}
	\includegraphics[width=\columnwidth]{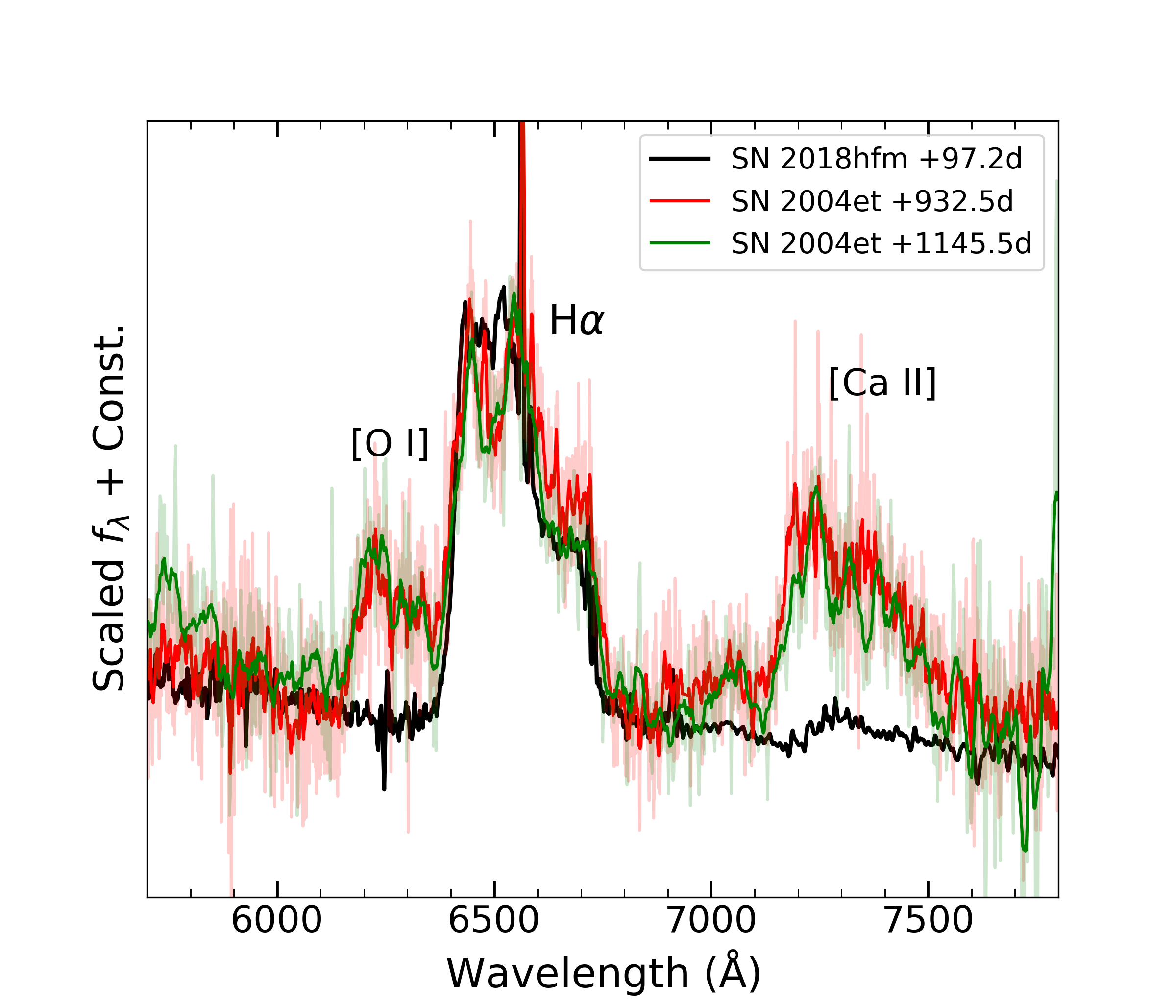}
        \caption{A late-time spectrum of SN 2018hfm compared with two very late-time spectra of SN 2004et \citep{Kotak2009}, showing the similarity in their broad and asymmetric H$\alpha$ profile. The spectra of SN 2004et are smoothed with a boxcar of suitable width.}
    \label{fig:compspec2}
\end{figure}

In Figure~\ref{fig:compspec2}, we compare a late-time spectrum of SN 2018hfm (+97.2\,d) with two very late-time spectra of SN 2004et (+932.5\,d and +1145.5\,d; \citealt{Kotak2009}). For the $\rm H\alpha$ emission, SN 2018hfm is quite similar to SN 2004et, both showing asymmetric box-like profiles. According to \citet{Jerkstrand2017_handbookSN}, a box-like profile 
is formed from a shell-like emission region. The maximum velocity at zero 
intensity (MVZI) of the profile corresponds to the outer boundary of the shell, and the maximum velocity of the flat top is related to the inner boundary. For SN 2014et and SN 2018hfm, this shell-like emission region is 
believed to have resulted from the collision of the outer ejecta with the 
CSM. Owing to dust coupled with the emission region, the profile is altered by extinction and scattering to show a red-blue asymmetry \citep{Bevan2016}. Compared with SN 2014et, SN 2018hfm shows no explicit signature of 
[\ion{O}{i}] $\lambda\lambda$6300, 6364, revealing that its progenitor has a very low oxygen abundance and hence a low main-sequence mass \citep{Woosley1995}. Additionally, the [\ion{Ca}{ii}] $\lambda\lambda$7291, 7323 emission of SN 2018hfm is still recognisable, meaning a relatively large flux ratio of $R =$ ([\ion{Ca}{ii}] $\lambda\lambda$7291, 7323)/([\ion{O}{i}] $\lambda\lambda$6300, 6364), which also points to a low-mass progenitor origin for SN 2018hfm \citep{Inserra2011}.
 
\subsection{Evolution of spectroscopic parameters}
\label{subsec:spec_para}
We measure the temperature and line velocity from the extinction- and redshift-corrected spectra of SN 2018hfm. The temperature is derived by applying blackbody fits to the spectra. For the velocity measurement, we first smooth the spectra when necessary, and then zoom in the absorption trough to judge the minimum by eye. We repeat the measurements by three times and average the results to obtain a final velocity value. The measurements are confined to spectra taken before +60\,d since explosion, as tail-phase spectra are dominated by emission features and flat continua which are likely contaminated by the host galaxy.\\

\begin{figure*}
	\includegraphics[width=2\columnwidth]{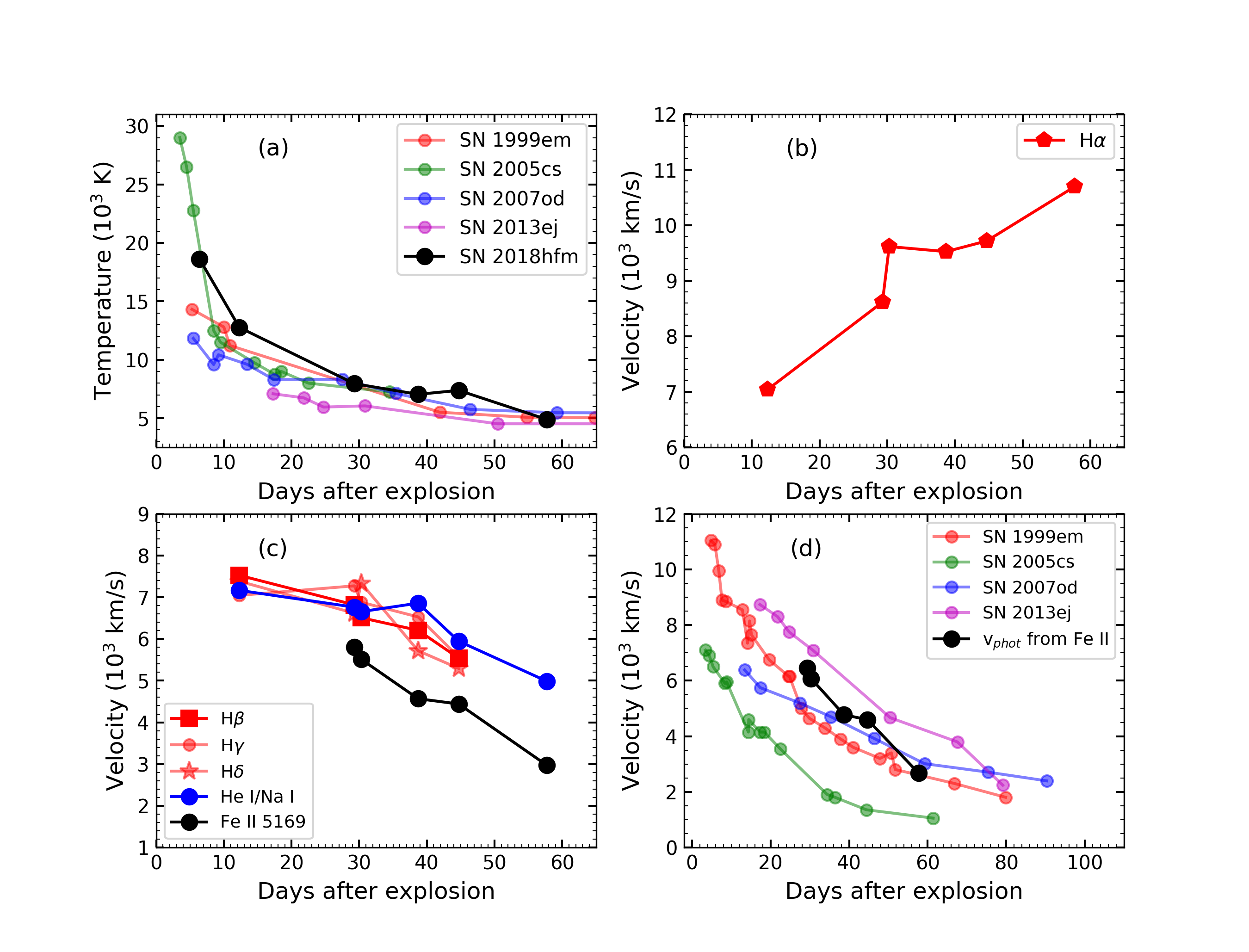}
    \caption{(a) Blackbody temperature evolution of SN 2018hfm compared with that of other SNe~II (SN 1999em, \citealt{Elmhamdi2003}; SN 2005cs, \citealt{Pastorello2006}; SN 2007od, \citealt{Inserra2011}; SN 2013ej, \citealt{Huang2015}).  (b) $\rm H\alpha$ velocity of SN 2018hfm inferred from minima of the absorption trough. (c) Line velocity of $\rm H\beta$, $\rm H\gamma$, $\rm H\delta$, \ion{He}{i} $\lambda$5876/\ion{Na}{i}, and \ion{Fe}{ii} $\lambda$5169 measured from spectra of SN 2018hfm. (d) Photoshperic velocity of SN 2018hfm inferred from $v_{\ion{Fe}{ii}}$ compared with that of other SNe~II \citep{Inserra2011, Takats2012, Huang2015}.}
    \label{fig:measure_spec}
\end{figure*}

As shown in Figure~\ref{fig:measure_spec}(a), the blackbody temperature of SN 2018hfm is found to be higher than that of SN 1999em and SN 2007od, but similar to SN 2005cs in the early phase. The high temperature of SN 2005cs drops quickly and reaches the same level as SN 1999em and SN 2007od 
at +10\,d, while SN 2018hfm remains hotter than them until one month after explosion. The temperature of SN 2013ej is much lower than that of SN 2018hfm for about two months.\\

Line velocities are shown in Figure~\ref{fig:measure_spec}(b) and (c). We 
present the $\rm H\alpha$ velocity in an individual panel to highlight its abnormal evolution. Normally, its velocity should decrease with time like that of other Balmer lines ($\rm H\beta$, $\rm H\gamma$, and $\rm H\delta$ shown in panel (c)), but it accelerates from $\sim 7000$\,km\,s$^{-1}$ at +10\,d to $\sim 11,000$km\,s$^{-1}$ shortly after the plateau phase. As discussed in the last paragraph of Sec.\ref{subsec:LC_para}, the deposited shock energy leads to normal SN evolution, in which line velocities decrease when the photosphere recedes into the deeper layers of the ejecta. However, photons created by CSI come from the outermost ejecta, whose 
velocity is rather large. These photons tend to fill in the absorption trough of P-Cygni profile and make the through shallower, which will result into a large-velocity measurement. When these photons gradually dominate in the spectrum, it is not unexpected that the H$\alpha$ reveals an abnormal acceleration. The reason $\rm H\beta$, $\rm H\gamma$, and $\rm H\delta$ are not influenced is that the CSI 
is still in its young phase ($< 1$\,yr) at this time, and the optical depth is so large that higher-order Balmer-series photons are converted into 
$\rm H\alpha$ efficiently, leading to a steep Balmer decrement \citep{CF94}.\\

As shown in Figure~\ref{fig:measure_spec}(c), \ion{He}{i} $\lambda$5876, probably blended with \ion{Na}{i} at later phases, evolves similarly to hydrogen, but \ion{Fe}{ii} $\lambda$5169 shows a lower velocity. It is consistent with the ``onion-ring'' structure of elements in progenitors of core-collapse SNe, with light elements lying in outer layers and heavy ones at smaller radii. Evolution of the photospheric velocity can be inferred from \ion{Fe}{ii} $\lambda$5169 through an empirical formula (see Eq.~1 
of \citealt{Takats2012} with parameters in their Table 2). As shown in Figure~\ref{fig:measure_spec}(d), SN 2018hfm, whose explosion energy is very low, exhibits photospheric velocity comparable to that of other SNe~II. 
This is because the ejecta mass of SN 2018hfm is also much lower than that of the comparison SNe.

\subsection{High-velocity features}
\begin{figure}
	\includegraphics[width=\columnwidth]{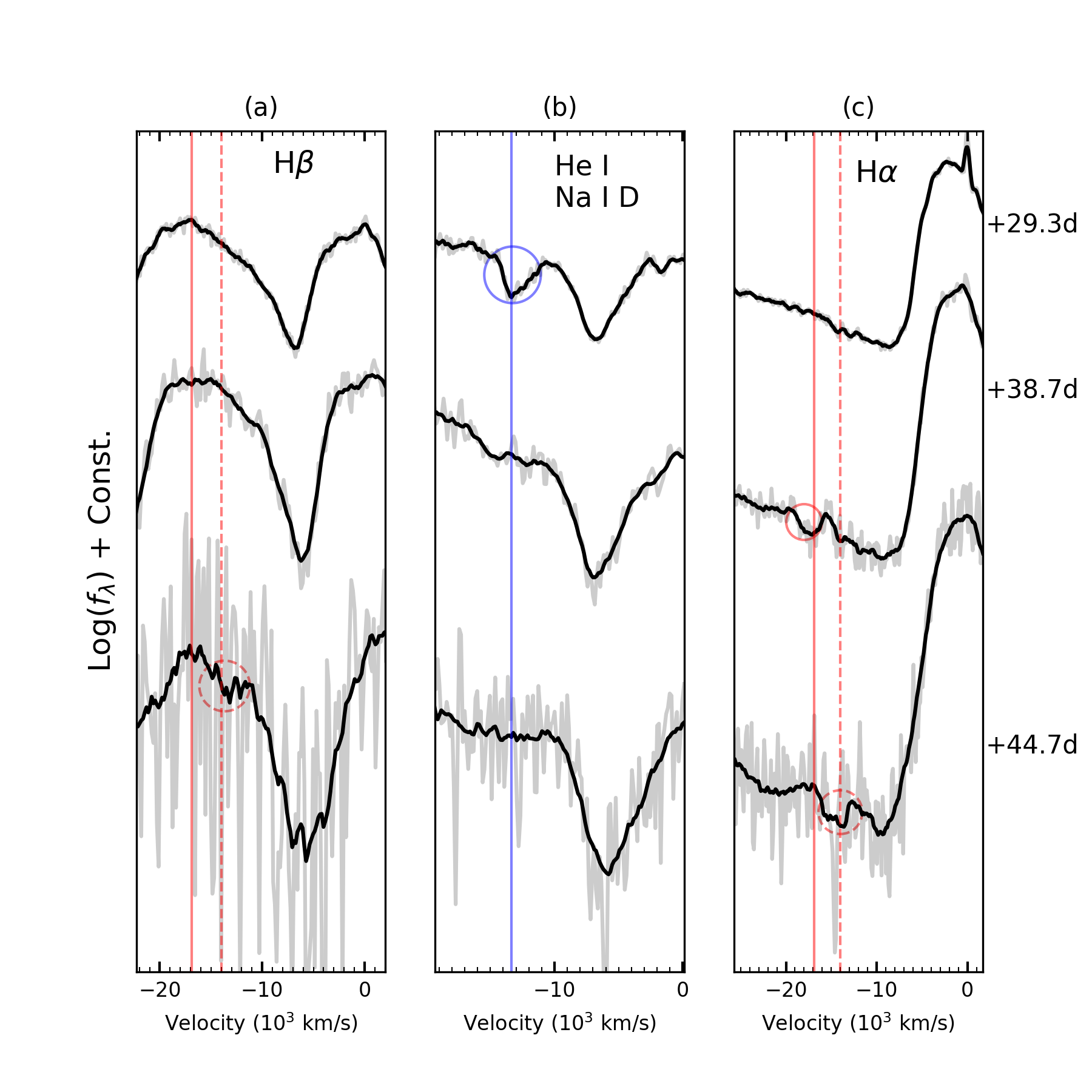}
    \caption{High velocity (HV) features of (a) $\rm H\beta$, (b) \ion{He}{i} $\lambda$5876, and (c) $\rm H\alpha$. Solid (dashed) red lines in panels (a) and (c) denote a velocity of $-16,900$ ($-14,000$) km\,s$^{-1}$, 
with solid (dashed)-edge circles highlighting the HV features. The blue line in panel (b) marks a velocity of $-13,340$\,km\,s$^{-1}$, with a blue 
circle denoting the HV feature of \ion{He}{i}. Phases of spectra are labelled at the right side of panel (c). Velocity is relative to the rest wavelength of each line.}
    \label{fig:HV}
\end{figure}

In the situation of CSI, X-rays, mainly from the reverse shock, ionise and excite the outer layers of SN ejecta to produce a small depression in the blue wing of the undisturbed absorption component of the P~Cygni profile \citep{Chugai2007, Blinnikov2017}. This small depression, or high-velocity feature, is observed in many SNe~II. For instance, it is reported by \citet{Gutierrez2017a} that 60\% of their sample exhibit this feature. 
We thus inspect spectra of SN 2018hfm to search for possible evidence of such a feature (see Figure~\ref{fig:HV}). As discussed in Section~\ref{subsec:compspec}, the blue notch on the left of \ion{He}{i} $\lambda$5876 disappears when the temperature decreases, which favours formation by high-velocity helium. As for hydrogen, if a blue notch in the $\rm H\alpha$ profile is indeed a high-velocity feature, then similar absorptions are expected in the profiles of other hydrogen lines \citep{Faran2014a, Gutierrez2017a}. Thus, only the small trough with a velocity of $-14,000$\,km\,s$^{-1}$ in the +44.7\,d spectrum can indeed be a high-velocity feature, while the notch in the +38.7\,d spectrum could be a signature of \ion{Si}{ii} $\lambda$6355 with a velocity of $-7634$\,km\,s$^{-1}$.\\

\subsection{Box-like emission of $\rm H\alpha$ and [\ion{Ca}{ii}] at late 
phase}
\label{subsec:H_CaII}
In this section, we focus on the evolution of $\rm H\alpha$ and [\ion{Ca}{ii}] emission lines emerging in the late phases (from +66.7\,d to +389.4\,d). We apply continuum subtraction and intensity normalisation to the $\rm H\alpha$ emission, and then we define and measure the following parameters from the emission-line profile --- the blue/red velocity at zero intensity (BVZI/RVZI), the blue velocity at half-maximum intensity (BVHM), and the velocity of the inner boundary of the shell-like emission region ($V_{\rm in}$) (see detailed description in Appendix \ref{sec:preprossing}). Results of these parameters are listed in Table~\ref{tab:Halpha_para}.\\

\begin{table*}
	\centering
	\caption{Values of parametars defined in Figure~\ref{fig:measure_Halpha} and Appendix \ref{sec:preprossing}.}
	\label{tab:Halpha_para}
	\begin{tabular}{cccccccc} 
		\hline
		Phase & BVHM & RVZI & $V_{\rm in}$ & BVZI ($V_{\rm out}$) & $V_{\rm in}/V_{\rm out}$ & right roof ratio & left roof  \\
		\hline
		 (d)  & (km\,s$^{-1}$)     & (km\,s$^{-1}$)   & km\,s$^{-1}$  & km\,s$^{-1}$    &      &      & (erg\,s$^{-1}$\,cm$^{-2}$\,\AA$^{-1}$)  \\
		\hline
		+66.7  & -7557.14 & 16,000 & -6500 & -13,500 & 0.48 & 0.50 & 1.279e-15  
 \\
		+75.6  & -7504.64 & 15,000 & -6400 & -11,000 & 0.58 & 0.50 & 7.640e-16  
 \\
		+89.6  & -7572.84 & 12,500 & -6766 & -10,600 & 0.64 & 0.42 & 3.261e-16  
 \\
		+97.2  & -7134.23 & 12,200 & -6245 & -10,600 & 0.59 & 0.46 & 6.975e-15  
\\
		+122.5 & -6699.28 & 10,300 & -5848 &  -9,000 & 0.65 & 0.60 & 3.879e-16  
 \\
		+143.4 & -7130.26 & 10,000 & -6100 & -10,600 & 0.58 & 0.33 & 3.928e-16  
 \\
		+182.3 & -6975.46 &  8,668 & -5480 &  -8,500 & 0.64 & 0.20 & 1.944e-16  
 \\
		+389.4 & -6398.80 &  8,400 & -5686 &  -7,877 & 0.72 & 0.08 & 6.617e-17  
 \\
		\hline
	\end{tabular}
\end{table*}

\begin{figure}
  \subfigure[]{
     \includegraphics[width=\columnwidth]{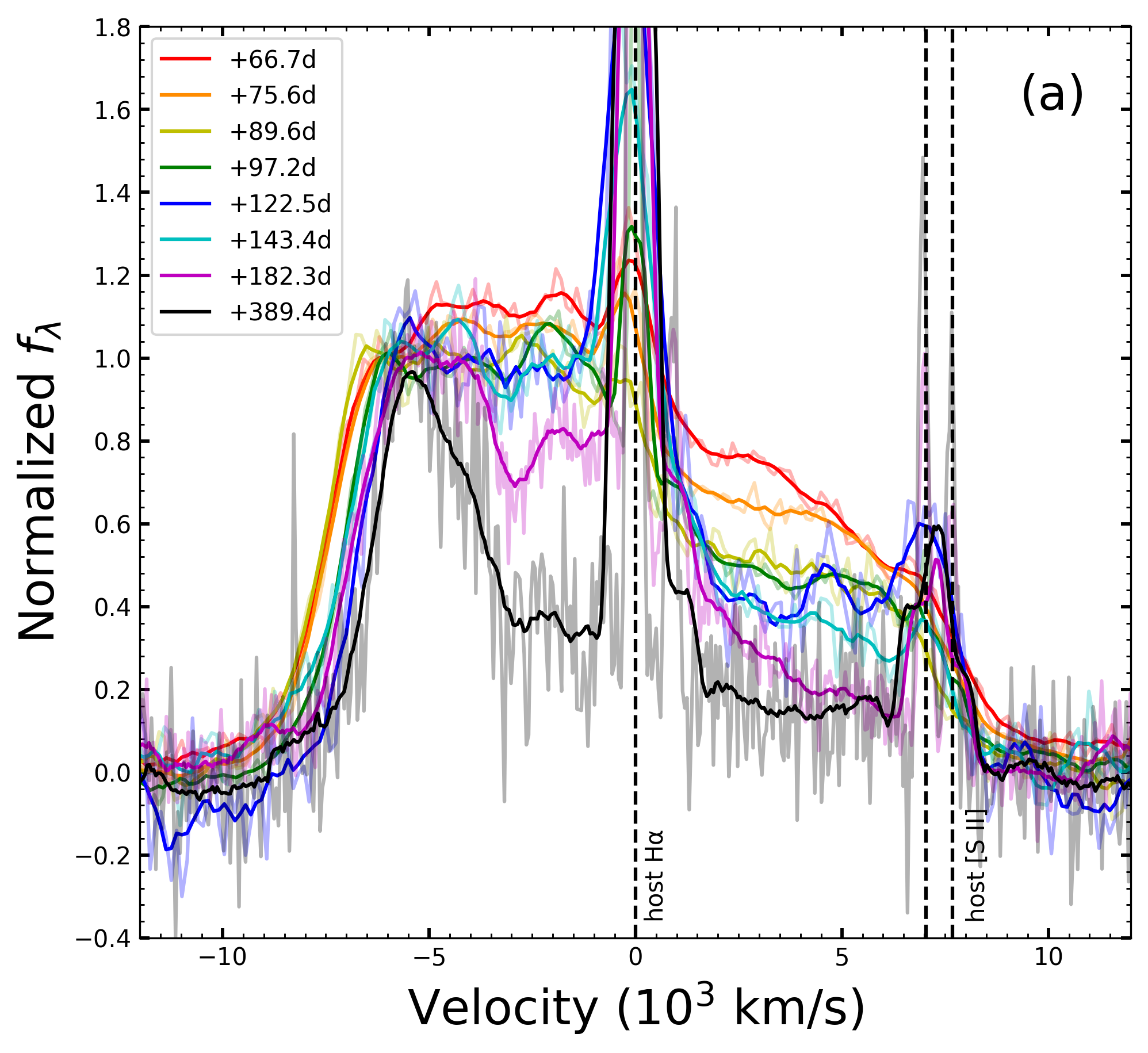}
  }
  \quad
  \subfigure[]{
     \includegraphics[width=\columnwidth]{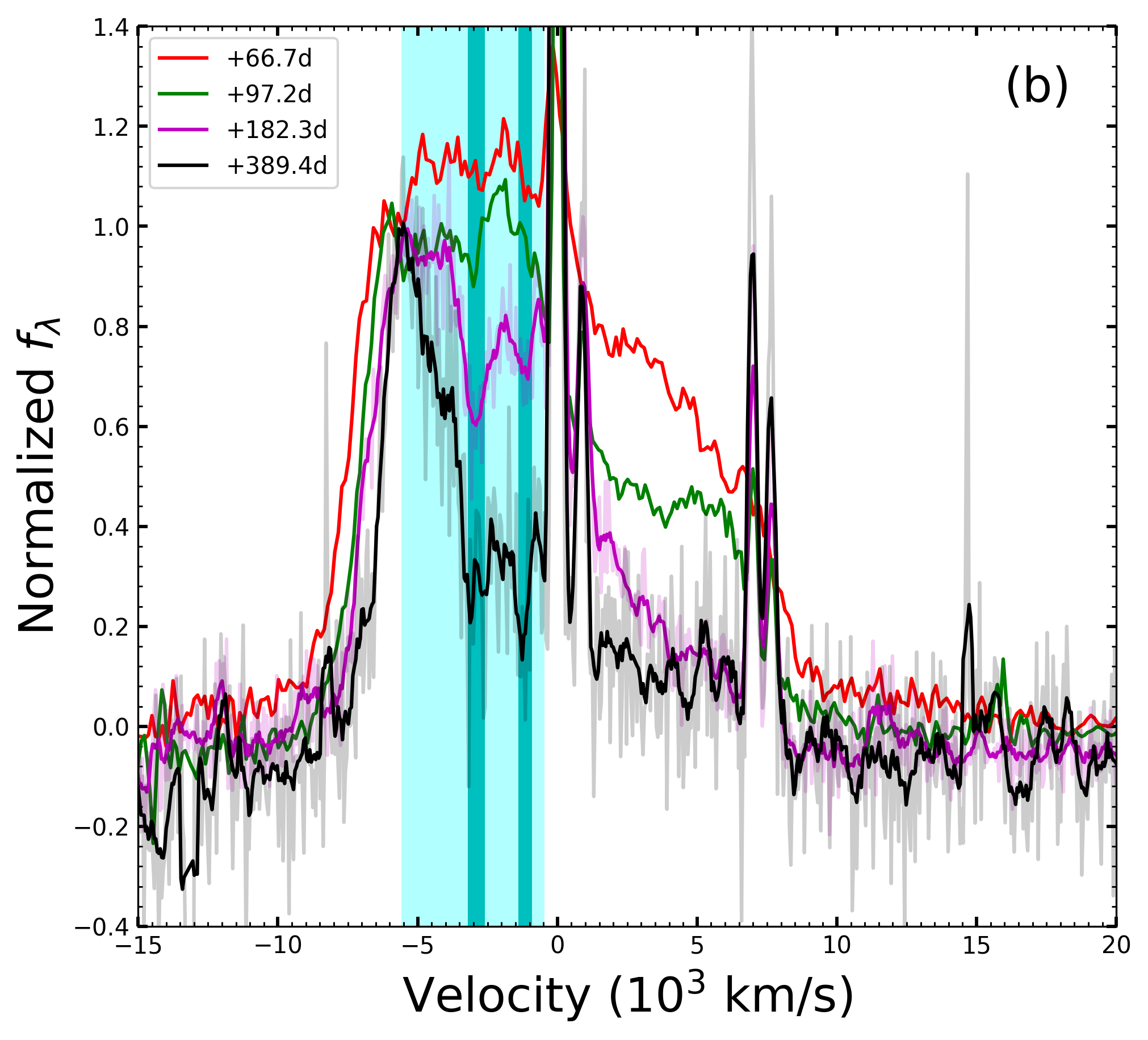}  
  }
  \caption{(a) Evolution of the box-like $\rm H\alpha$ emission profile. Phase is coded with colour. All of the spectra are smoothed by a boxcar of suitable width, with the smoothed spectrum superimposed on the original 
one. Dashed lines denote narrow emission lines from host \ion{H}{ii} region. (b) Four of the same spectra, to show the evolution of the CDS component in detail. The broad cyan region denotes the blue side of the CDS component. Two narrow dark-cyan regions mark the troughs in the profiles.}
  \label{fig:Halpha}
\end{figure}

From Figure~\ref{fig:Halpha}(a) or from the values of BVHM/BVZI listed in Table~\ref{tab:Halpha_para}, one can find that the $\rm H\alpha$ box-like profile is very broad, with a velocity of $\sim 10,000$\,km\,s$^{-1}$, but the line width decreases with time. The broad line profile suggests 
that it is not from radioactive decay because radioactive decay of heavy elements usually occurs deep inside the SN ejecta \citep{Chugai1990}; also, the decreasing line width excludes the main energy contribution being from a pulsar, because acceleration of a pulsar bubble will lead to broadening of the emission lines \citep{CF92}. According to \citet{CF94}, CSI can naturally explain the above phenomena. Energy from CSI heats/ionises the outer-layer ejecta and the cold dense shell (CDS), making them emit recombined $\rm H\alpha$ photons and forming box-like emission profiles in 
the spectra. Since the kinetic energy is consumed, the velocity of the ejecta decreases with time, and the emission line thus becomes narrow. From 
Figure~\ref{fig:Halpha}(a), one can also see that the red-blue asymmetry increases with time, indicating that more dust is formed in the emission region \citep{Bevan2019}.\\

Because both the ionised ejecta and the CDS emit $\rm H\alpha$, to show evolution of the CDS component in detail we select four spectra (+66.7\,d, 
+97.2\,d, +182.3\,d, and +389.4\,d) in Figure~\ref{fig:Halpha}(a) and replot them in Figure~\ref{fig:Halpha}(b). The width of the CDS component is 
only $\sim 5000$\,km\,s$^{-1}$, consistent with the prediction that emission lines from the CDS have intermediate width (a few $10^3$\,km\,s$^{-1}$; \citealt{Smith2017_handbookSN}). The top of the profile is not very flat and shows many small-scale structures, within which two troughs develop with time. This means clumping exists in the emission region and the clumping gradually becomes severe \citep{Jerkstrand2017_handbookSN}, probably owing to cooling and increasing of thermal instability \citep{Inserra2011}. Here, however, we cannot determine whether this clumping occurs in the ionised ejecta, the CDS, or both; this question will be answered in Section~\ref{subsec:dust_formation}.\\

\begin{figure}
  \includegraphics[width=\columnwidth]{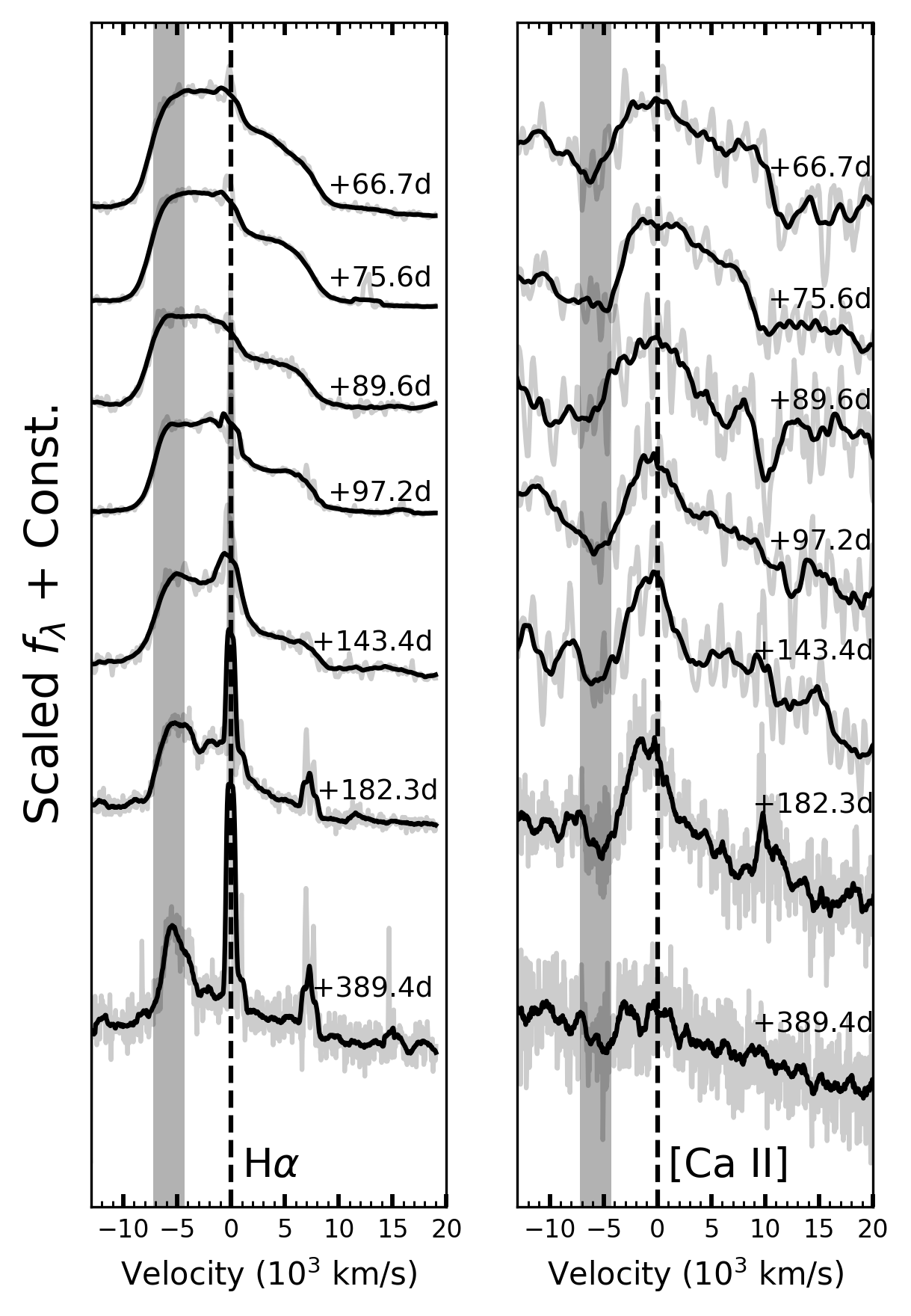}
  \caption{Comparison between evolution of $\rm H\alpha$ emission (\textit{left panel}) and [\ion{Ca}{ii}] $\lambda\lambda$7291, 7323 (\textit{right panel}). The $\rm H\alpha$ velocity is relative to 6563\,\AA\ and the [\ion{Ca}{ii}] velocity is relative to 7307\,\AA. Vertical dashed lines mark zero velocity and shaded regions in both panels denote the velocity between $-7200$\,km\,s$^{-1}$ and $-4300$\,km\,s$^{-1}$. This shaded region covers the blue minimum of [\ion{Ca}{ii}], but not for $\rm H\alpha$, which is located at larger velocity beyond this region. Note that the continuum has not been removed.}
  \label{fig:CaII}
\end{figure}

Another observed emission line in late-phase spectra is [\ion{Ca}{ii}] $\lambda\lambda$7291, 7323. As shown in Figure~\ref{fig:CaII}, [\ion{Ca}{ii}] also exhibits a box-like shape and broad width, but its BVZI ($-7200$\,km\,s$^{-1}$ to $-4300$\,km\,s$^{-1}$) is lower than that of $\rm H\alpha$. This intermediate width indicates that [\ion{Ca}{ii}] emission is very likely from the CDS, as suggested by \citet{CF94}, who argued that it should arise almost exclusively in the CDS. Similarly, [\ion{Ca}{ii}] reveals increasing red-blue asymmetry, which can be attributed to dust formation. Compared to $\rm H\alpha$, however, the red side of [\ion{Ca}{ii}] emission extends to a larger velocity than its blue side. This could result from very severe scattering, or there exists emission of other elements that we have not identified.\\

\section{Discussion}
\label{sec:discussion}
\subsection{When does the CSI begin?}

For SN 2018hfm, interaction between the ejecta and CSM is confirmed by several pieces of evidence, including the small bump at the end of the plateau in the light curves, high-velocity features of hydrogen and helium emerging in photospheric spectra, and the box-like $\rm H\alpha$ emission-line profile seen in late-phase spectra. However, when does the interaction actually begin?\\

The bell-shaped $\rm H\alpha$ profile in the +57.7\,d spectrum can be decomposed into a shallow-absorption P~Cygni profile and a box-like shape, suggesting that interaction must occur before this time. The small bump emerging in the $V/g$-band light curve suggests the CSI should occur at least by $\sim 40$\,d after explosion. However, as shown in Figure~\ref{fig:measure_spec}(b), $\rm H\alpha$ shows abnormal acceleration owing to the influence of CSI. This acceleration begins before +30\,d, indicating that the interaction should exist at an even earlier phase. Considering that the outer-layer ejecta have a velocity of $\sim 10,000$\,km\,s$^{-1}$, and assuming the interaction occurs at $\sim 30$\,d after explosion, we estimate that the CSM is located at a radius of $\sim 2.6 \times 10^{15}$\,cm ($\sim 170$ au) from the progenitor. Adopting a wind velocity of $\sim 10$\,km\,s$^{-1}$, we find that the CSM was produced at $t \approx 80$\,yr before the SN explosion.

\subsection{Progenitor scenario of SN 2018hfm}
Progenitors of many SNe~II have been identified as red supergiant (RSG) stars in pre-explosion images (\citealt{Smartt2009}, and references therein). For SN 2018hfm, despite no images before explosion being found, we can discuss its possible progenitor scenario through characteristics of the SN evolution.\\

For the light-curve morphology, SN 2018hfm has a luminous peak, a large plateau slope, and a short plateau duration. From modelling the bolometric light curve, we find that SN 2018hfm has a relatively low explosion energy of $\sim10^{50}$\,erg. Considering that the main energy source of the tail-phase light curve is likely from CSI (see Sec.\ref{subsec:color_curve}), the mass of $^{56}$Ni is expected to be very small, or possibly there is no contribution of $^{56}$Ni at all. These features are reminiscent of the low-energy SN progenitor study of \citet{Lisakov2018}. They evolve a single star with an initial mass of 27\,$\rm M_{\odot}$ to the pre-SN phase and explode the star with a low energy. They find that this high-mass progenitor tends to retain small mass in its hydrogen envelope before explosion owing to great mass loss during the RSG phase. This low-mass envelope leads to a light curve with a relatively bright peak, fast post-peak decline, and short plateau duration, which is quite similar to the light curve of SN 2018hfm. And in their low-energy explosion model, the entire CO core falls back, and hence no $\rm ^{56}Ni$ is expelled outside. This is a possible explanation if the tail luminosity of SN 2018hfm is completely supplied by CSI. Moreover, \citet{Lisakov2018} predict in their model that the SN shows a bluer colour at early phases than normal SNe~II. They attribute this to the large radius ($>800\,\rm R_{\odot}$) of the  pre-SN progenitor, which impacts the cooling from expansion. We observe the bluer colours and we infer that the progenitor of SN 2018hfm has an extended radius ($>1000\,\rm R_{\odot}$). However, the large flux ratio between [\ion{Ca}{ii}] and [\ion{O}{i}] seen in the late-time spectra suggests that the progenitor mass of SN 2018hfm is not as large as 27\,$\rm M_{\odot}$.\\

\citet{Reguitti2021} describe some observational similarities between low-energy SNe~II and electron-capture SNe (ECSNe; \citealt{Tominage2013, Moriya2014}), e.g., low explosion energy and little contribution of $^{56}$Ni at late phase. ECSNe are believed to be the outcome of super-asymptotic giant branch (super-AGB) stars. According to \citet{Pumo2009}, super-AGB stars are those which have an inert core (composed of Ne and O) and an envelope of burning helium and hydrogen. The upper limit of the initial mass for a super-AGB star is about $10 \sim 11\,\rm M_{\odot}$, which is at the small end of the SNe II progenitor mass range. These super-AGB stars usually experience thermal pulses, so that some mass of the envelope is ejected into the space; meanwhile, some mass is thrown onto the core. When the core mass goes beyond the Chandrasekhar limit ($\sim 1.37\,\rm M_{\odot}$), electron capture reactions occur and finally lead to a core-collapse explosion. However, compared with normal iron-core-collapse SNe, this class of SNe tend to have relatively low explosion energy ($\sim 10^{50}$\,erg). Some features of SN 2018hfm are similar to those of ECSNe, such as the low explosion energy of $\sim 10^{50}$\,erg, very low mass of $^{56}$Ni, possibly low progenitor mass inferred from the large flux ratio between [\ion{Ca}{ii}] and [\ion{O}{i}], and mass-loss history verified by the CSI signature. However, it is difficult to declare that SN 2018hfm is an ECSN simply based on these plausible lines of evidence.\\

\subsection{Dust formation in the ejecta}
\label{subsec:dust_formation}
To quantify how much dust is formed in SN 2018hfm, we use the tool \textsc{damocles} \citep{2018ascl.soft07023B, Bevan2016}, which is a Monte Carlo code modelling the influence of dust attenuation/scattering on optical/near-infrared emission lines. We construct a very simple configuration, assuming a shell-like emission region with homologous expansion, with velocity of the inner boundary ($V_{\rm in}$) and outer boundary ($V_{\rm out}$) set to the values listed in Table~\ref{tab:Halpha_para}. Between $V_{\rm in}$ and $V_{\rm out}$, the density of emitting material (here hydrogen) follows a smooth radial power-law distribution with an index of 5 (i.e., $\rho \propto V^{-5}$). Assuming that the hydrogen is ionised completely, the emissivity ($i$) is proportional to the square of the density: $i \propto V^{-10}$.\\

In addition, this emission region is assumed to be optically thin and suffer only dust absorption, without any scattering. The dust is set to be coupled with the emitting material, so that its density follows the same power-law distribution ($\rho_{\rm dust} \propto V^{-5}$). The composition 
of the dust is presumed as BE amorphous carbon \citep{Zubko1996}, with a grain radius of 0.01\,$\mu$m . We vary the dust mass so that the model can fit the observations well. Note that this configuration is only applied 
to the ionised ejecta; we do not model the emission from the CDS in this work, because the line profile of the CDS component reveals possible severe scattering. Thus, it is difficult for us to determine the dust mass coupled in the CDS through the simple model. As the CDS is usually very thin and has a very low mass compared with the ejecta, we expect that a large amount of dust was not formed in the CDS. In the following discussion, any dust in the CDS is omitted.\\

\begin{figure*}
  \includegraphics[width=1.5\columnwidth]{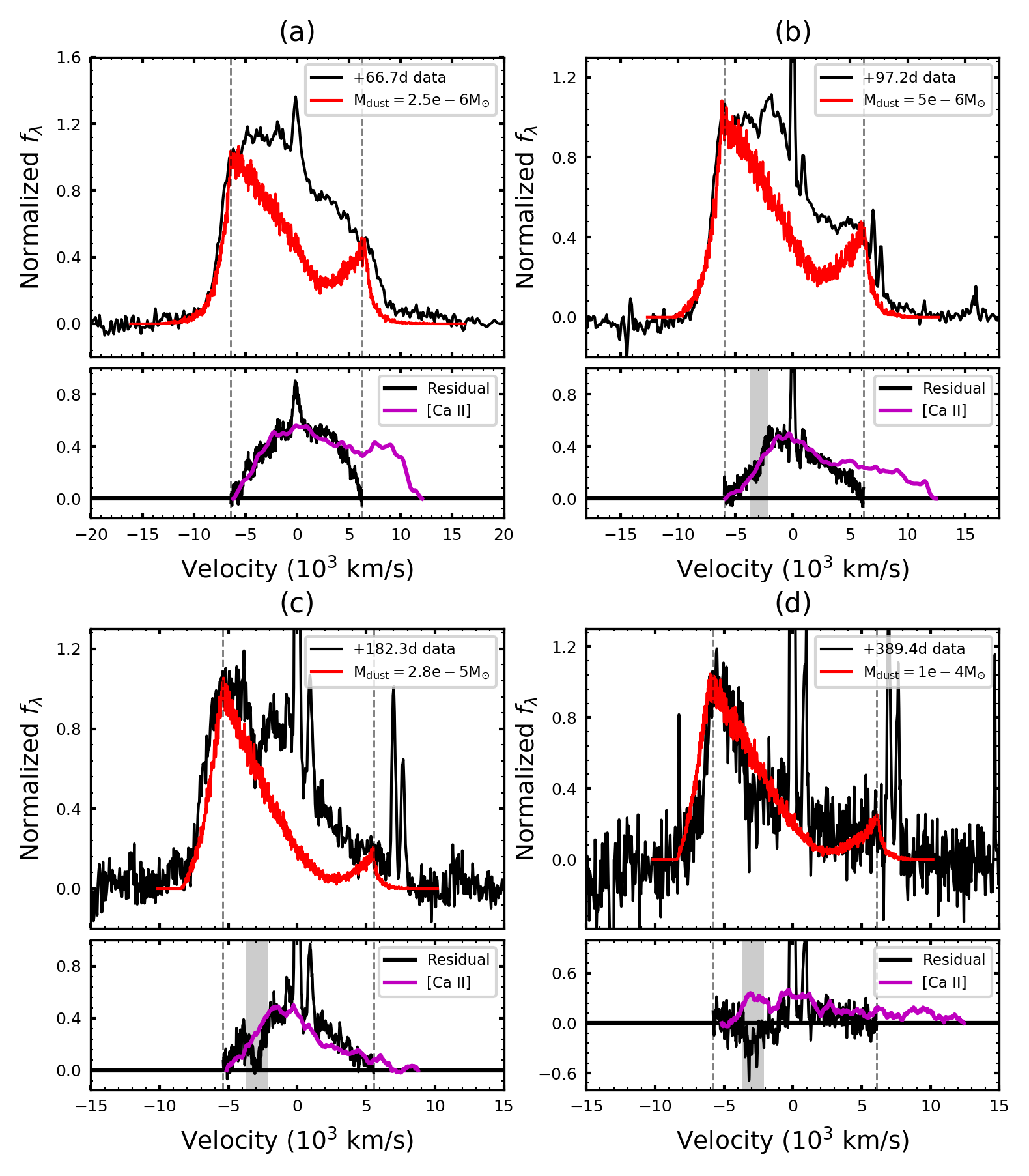}
  \caption{Box-like $\rm H\alpha$ emission lines with the dust model from 
\citet{Bevan2016}. We show data and models at four different epochs: (a) +66.7\,d, (b) +97.2\,d, (c) +182.3\,d, and (d) +389.4\,d. Black lines in all of the upper panels represent the observed data, while red lines show 
the model. All models  are set to the same parameters except the dust mass as denoted in the label. Residuals of the data after subtracting the model, between two vertical dashed lines, are shown in the lower panels in black lines. The magenta lines superimposed on the residuals are profiles 
of [\ion{Ca}{ii}] $\lambda\lambda$7291, 7323 emission at the same epoch. Grey shaded regions in the lower panels denote the trough in the residual 
data.}
  \label{fig:dust_model}
\end{figure*}

We simulate the $\rm H\alpha$ emission at four epochs (+66.7\,d, +97.2\,d, +182.3\,d, and +389.4\,d), which cover the time from the beginning of tail evolution to very late phases. The four emission lines, together with 
the best-fit model, are presented in the upper panels of Figure~\ref{fig:dust_model}. For convenience of discussion, we split the $\rm H\alpha$ line profile into three parts according to their left and right shoulders (see vertical dashed lines in Figure~\ref{fig:dust_model}). For the left part, our model reproduces the data very well, meaning that the assumption of a power-law distribution for the emitting materials is reasonable. In fact, it is exactly the case that outer layers of CCSN progenitors can be mimicked by a steep power law \citep{CF94}.\\

For the right part, our model underestimates the data because scattering actually exists in the emission region. Large discrepancies exist in the middle part, where the residuals are shown in the lower panels of Figure~\ref{fig:dust_model}. These residuals are believed to be caused by the CDS. To verify this idea, we extract [\ion{Ca}{ii}] $\lambda\lambda$7291, 7323 emission lines in the same spectra and superimpose them on the residuals after suitable rescaling. One can see that the blue sides of the residuals are consistent with those of [\ion{Ca}{ii}] at all of the four selected epochs, which convincingly favours that they are from the same emission region, namely the CDS, as discussed in Section~\ref{subsec:H_CaII}. The fact that the red side of [\ion{Ca}{ii}] extends to a larger velocity 
than the residuals is probably due to blending with other emission lines. 
In Section~\ref{subsec:H_CaII}, we attribute the small troughs on the top 
of the $\rm H\alpha$ emission profile to clumping in the emission region, 
but we cannot determine whether the clumping is located in the ionised ejecta or in the CDS. From comparison with the [\ion{Ca}{ii}] line profiles, we find the trough exists in the residuals but not in [\ion{Ca}{ii}]. Therefore we conclude that the troughs should come from the ionised ejecta.\\

\begin{figure}
  \includegraphics[width=\columnwidth]{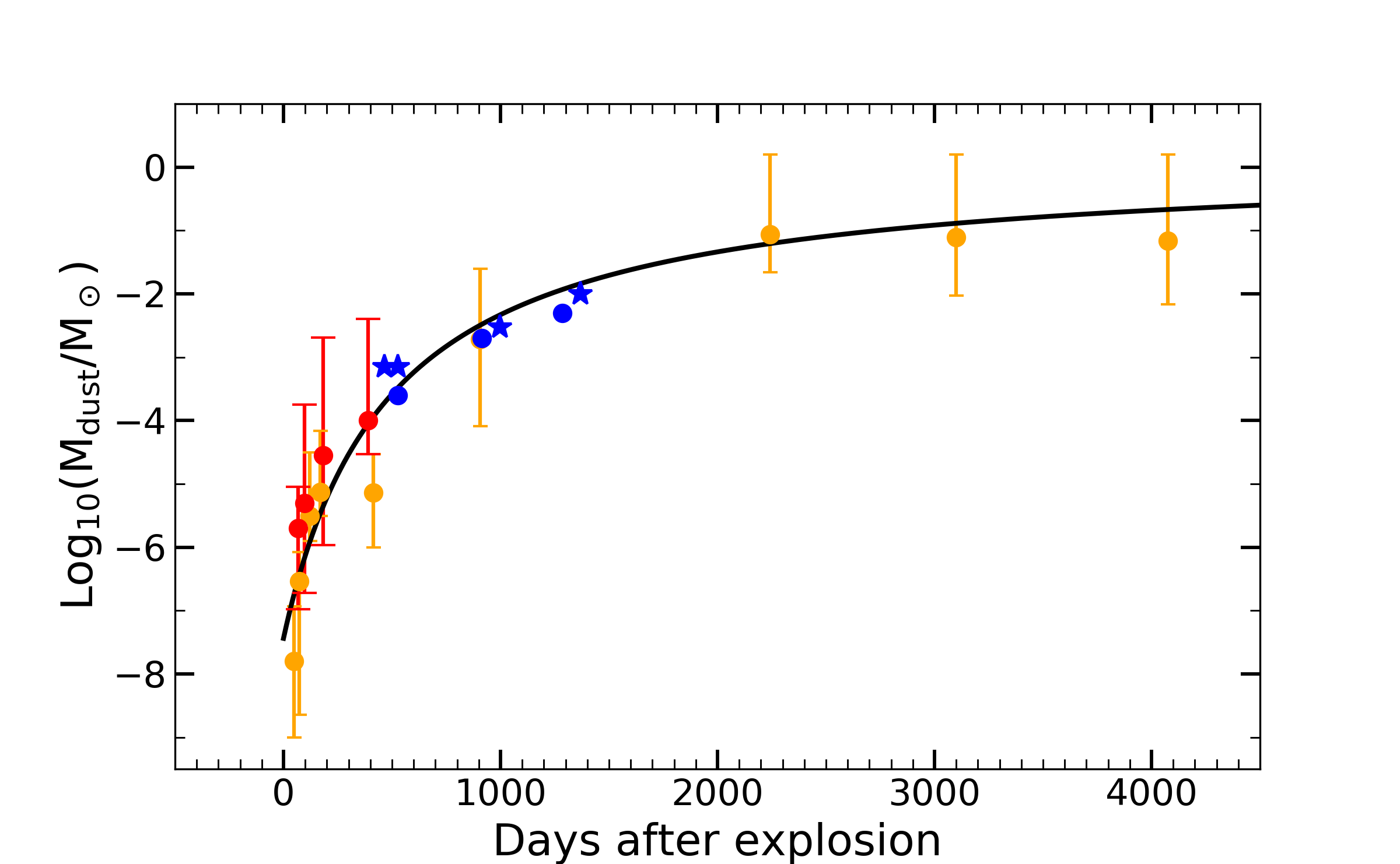}
  \caption{Dust-mass evolution of SN 2018hfm (red circles; this paper), SN 2005ip (orange cicles; \citealt{Bevan2019}), and SN 2010jl (blue circles and blue stars; \citealt{Bevan2020}). Circles represent dust mass estimates through the red-blue asymmetry of optical emission lines, while stars represent estimates through fitting the infrared SEDs. The error bars are the 16th and 84th quartiles of the 1D dust-mass posterior distributions (see text). The black line shows the best-fit curve with the form of Eq.~\ref{eq:Bevan_curve} proposed by \citet{Bevan2019}.}
  \label{fig:dust_formation_line}
\end{figure}

From the models described above, we find that dust is formed continuously as time elapses, from $2.5\times10^{-6}\,\rm M_{\odot}$ on +66.7\,d to 
$1\times10^{-4}\,\rm M_{\odot}$ on +389.4\,d. To examine the reliability of these dust-mass estimates, we use a Bayesian approach characterised by 
the application of an affine invariant Markov Chain Monte Carlo ensemble sampler to \textsc{damocles}, to produce marginalised one-dimensional (1D) posterior probability distributions of our input parameters. This method was first used for SN 1987A by \citet{Bevan2018}. We take the 16th and 84th quartiles of the 1D dust-mass posterior distributions as lower and upper limits at each epoch, and find that our dust-mass estimation of SN 2018hfm at all four epochs coincides with the uncertainty ranges between the two limits. As shown in Figure~\ref{fig:dust_formation_line}, even though the uncertainty ranges are somewhat large (across $\sim 2$ orders of magnitude, possibly owing to the low signal-to-noise ratio of the spectra), it is clear that dust mass is newly formed throughout an interval of $\sim300$\,d, increasing from a relatively low mass range ($10^{-7}$--$10^{-5}\,\rm M_{\odot}$) to a high mass range ($10^{-4}$--$10^{-3}\,\rm M_{\odot}$). Dust formation has been also observed in several other SNe~II, such as SN 1987A \citep{Bevan2016}, SN 2005ip \citep{Bevan2019}, and SN 2010jl \citep{Bevan2020}. \citet{Bevan2020} find that the dust-formation rate of these three SNe obeys a best-fit curve with the form
\begin{equation}
    -\log_{10}\left(\frac{M_{\rm dust}}{\rm M_{\odot}}\right) = c\left(a\frac{t}{\rm days}+b\right)^{-1} + d,
    \label{eq:Bevan_curve}
\end{equation}
where $a=0.0076$, $b=3.66$, $c=27.7$, and $d=-0.13$. We present this curve in Figure~\ref{fig:dust_formation_line}, overplotted with the dust mass measured from SN 2018hfm, SN 2005ip, and SN 2010jl. One can see that the dust-formation rate of SN 2018hfm also follows this curve, indicating that more dust is expected to be produced in this SN at later phases.

\section{Conclusions}
\label{sec:conclusion}
We present extensive photometric data for SN 2018hfm, covering the rise, plateau-like phase, transitional stage, and tail phase, from which we estimate an explosion date of MJD $= 58395.2 \pm 5.3$. The $V$-band light curve has a peak of $-18.69 \pm 0.64$\,mag, followed by a very rapid decline with rate of $\rm s_2=4.42\pm0.13$\,mag\,(100\,d)$^{-1}$. After about 50 days, the $V$-band light curve abruptly drops by $\sim 3$\,mag and then enters the tail phase. From the reconstructed bolometric light curve, we find that the ejecta mass of SN 2018hfm is very low, only $\sim 1.3\,\rm M_{\odot}$. The low tail luminosity indicates a small mass of $^{56}$Ni ($<0.015\,\rm M_{\odot}$) produced in the explosion, consistent with the low explosion energy ($0.3 \times 10^{51}$\,erg).\\

Extensive optical spectra of this SN, spanning from +6.4\,d to +389.4\,d after explosion, are shown. At very early phases, they are characterised by a blue featureless continuum. Through comparisons with other SNe~II, metal lines, such as \ion{Ca}{ii}, \ion{O}{i}, and \ion{Fe}{ii}, are identified in the photospheric spectra. During very late phases, spectra of SN 
2018hfm exhibit asymmetric box-like emission-line profiles of $\rm H\alpha$ and [\ion{Ca}{ii}] $\lambda\lambda$7291, 7323, which are tightly related to circumstellar interaction and dust formation. Through modelling the 
line profiles, we estimate the dust mass produced in the ionised ejecta, finding that the dust increases from $\sim10^{-6}\,\rm M_{\odot}$ at +66.7\,d to $10^{-4}$--$10^{-3}\,\rm M_{\odot}$ at +389.4\,d. This dust-formation rate follows a curve proposed by \citet{Bevan2019}, similar to that of SN 1987A, SN 2005ip, and SN 2010jl.\\

Based on the observational features and the parameters inferred from the SN data, such as a low mass of $^{56}$Ni, a low explosion energy, and the 
mass-loss history, we discuss possible progenitor scenarios of SN 2018hfm. We find that these features link the SN to a possible electron-capture explosion, but an iron-core-collapse SN exploded with low energy can also form most of the observational characteristics.\\

\section*{Acknowledgements}
The authors are grateful to some colleagues in the SN group or in THCA for useful suggestions on this paper. We thank Dr. C. Hao, L. Hu, A. Singh and P. Chen for their helps with our work in different aspects. The authors acknowledge support for observations from the staffs of XLT, TNT, LJT, APO, Lick, and Keck Observatories. The operations of XLT and 80cm Tsinghua-NAOC telescope were partially supported by the Open Project Program of the Key Laboratory of Optical Astronomy, National Astronomical Observatories, Chinese Academy of Sciences. Funding for the LJT has been provided by the Chinese Academy of Sciences and the People's Government of Yunnan Province. The LJT is jointly operated and administrated by Yunnan Observatories and Center for Astronomical Mega-Science, CAS. The W. M. Keck Observatory is operated as a scientific partnership among the California Institute of Technology, the University of California and NASA; the observatory was made possible by the generous 
financial support of the W. M. Keck Foundation. A major upgrade of the Kast spectrograph on the Shane 3\,m telescope at Lick Observatory was made possible through generous gifts from William and Marina Kast as well as the Heising-Simons Foundation. Research at Lick Observatory is partially supported by a generous gift from Google. We thank undergraduate students Jackson Sipple, Kevin Tang, Jeremy Wayland, and Keto Zhang for obtaining images with the 1\,m Nickel telescope.\\

This work is supported by National Natural Science Foundation of China (NSFC grants 12033003, 11633002, and 11761141001) and the National Program on Key Research and Development Project (grant 2016YFA0400803). This work 
is partially supported by the Scholar Program of Beijing Academy of Science and Technology (DZ: BS202002). This work is also supported by the Strategic Priority Research Program of the Chinese Academy of Sciences (grant XDB23040100). Y.-Z. Cai is funded by China Postdoctoral Science Foundation (grant no. 2021M691821). J.J.Z. is supported by the NSFC (grants 
11773067 and 11403096), the Key Research Program of the CAS (grant KJZD-EW-M06), the Youth Innovation Promotion Association of the CAS (grant 2018081), and the CAS ``Light of West China'' Program. T.M.Z. is supported by 
the NSFC (grant 11203034). Support for A.V.F.'s group at U.C. Berkeley was provided by the TABASGO Foundation, the Christopher R. Redlich Fund, and the Miller Institute for Basic Research in Science (where A.V.F. is a Senior Miller Fellow).

\section*{Data Availability}
Photometric data for SN 2018hfm are presented in Table~\ref{tab:ASNphoto}, Table~\ref{tab:NickelTphoto} and Table~\ref{tab:TNTphoto}. Spectroscopic data are available on reasonable request to the corresponding authors.

\bibliographystyle{mnras}
\bibliography{SN2018hfm} 



\appendix
\section{Preprocessing on box-like $\rm H\alpha$ emission}
\label{sec:preprossing}

\begin{figure*}
	\includegraphics[width=2\columnwidth]{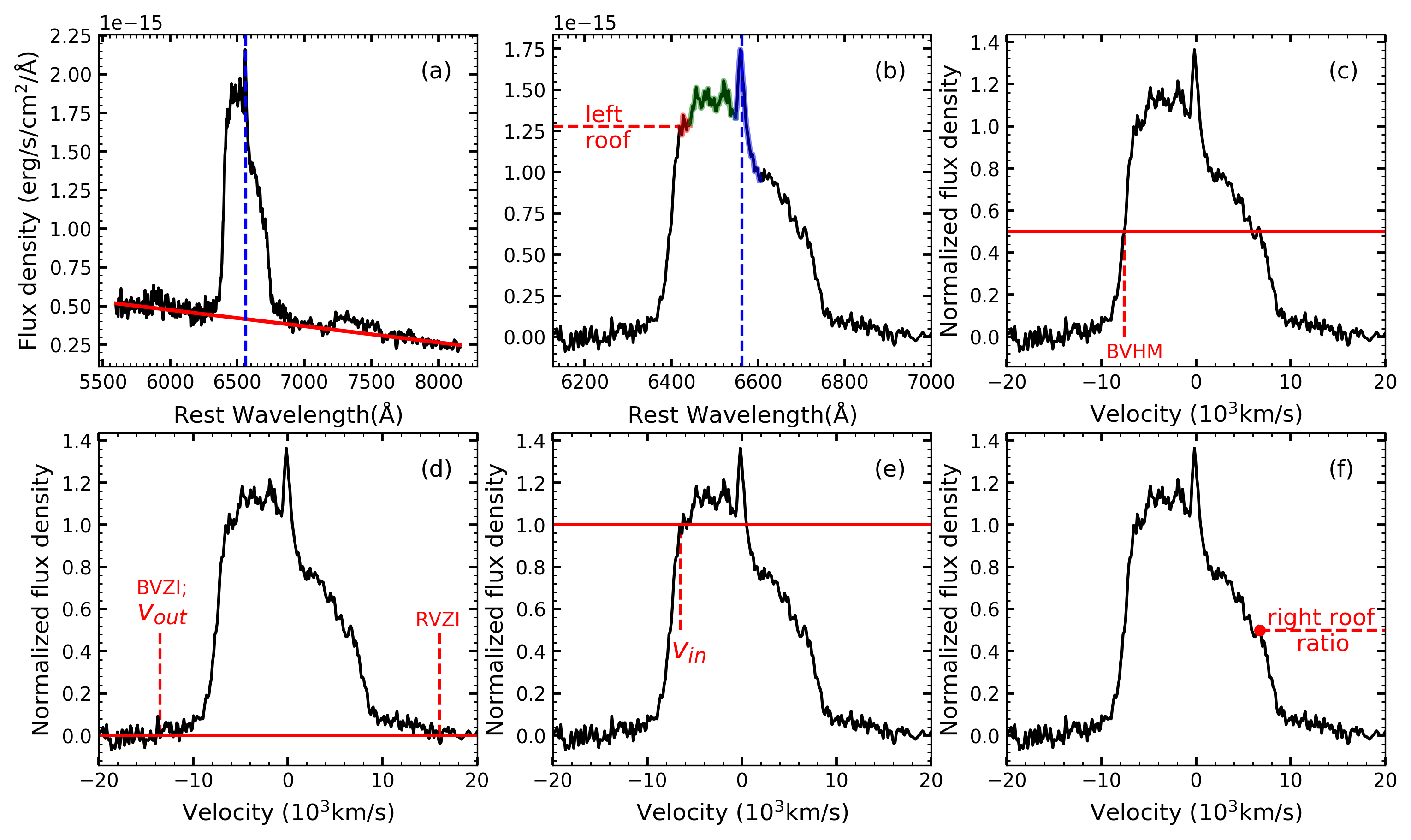}
    \caption{Demonstration of preprocessing the box-like $\rm H\alpha$ emission. For all six panels, black lines are $\rm H\alpha$ emission in the 
spectrum taken +66.7\,d after explosion. (a) The red line is fit to the continuum. The dashed line denotes the location of 6563\,\AA. (b) The red dashed line defines the ``left roof.'' Three components of the profile are highlighted by three colour (red, green, blue). The blue dashed line denotes the location of 6563\,\AA. (c) Definition of BVHM. (d) Definition of BVZI (equal to $V_{\rm out}$) and RVZI. (e) Definition of $V_{\rm in}$. 
(f) Definition of ``right roof ratio.''}
    \label{fig:measure_Halpha}
\end{figure*}

Taking Figure~\ref{fig:measure_Halpha} as an example, we describe the preprocessing performed on the box-like $\rm H\alpha$ emission line. First, a straight line was fit to the continuum, as shown in panel (a); it was then subtracted from the profile. We defined a ``left roof'' on the height 
of the left top edge (see panel(b)), and divided the profile by this value to normalise the profile. We determined the blue velocity at half maximum (BVHM), blue/red velocity at zero intensity (B/RVZI), inner boundary velocity ($V_{\rm in}$), and ``right roof ratio,'' shown in the following panels.\\

Note that in panel (b) of Figure~\ref{fig:measure_Halpha}, three lines (red, green, and blue) are superimposed on the top of the profile. The blue 
component is likely from an \ion{H}{ii} region in the host galaxy. The green component is believed to be from the cold dense shell (CDS). The red line denotes the left shoulder of the main box-like profile, which is from recombination of the ionised outer ejecta.\\

The box-like profile is formed from a shell-like emission region. Velocities of its inner boundary ($V_{\rm in}$) and outer boundary ($V_{\rm out}$) are tightly related to the line profile, as shown in panels (d) and (e). The ratio of the right to the left height of the profile is defined as 
``right roof ratio'' in panel (f), which reflects asymmetry of the profile and is related to dust properties.\\

All of the values of the parameters defined above are listed in Table~\ref{tab:Halpha_para}.

\bsp	
\label{lastpage}
\end{document}